\documentclass[11pt]{article}
\pdfoutput=1
\usepackage{jheppub}
\usepackage[T1]{fontenc}
\usepackage{cancel,tensor,enumitem,fix-cm}
\usepackage{epsfig}
\usepackage{graphicx}
\usepackage[english]{babel}
\usepackage{amsmath,amssymb}
\usepackage{dsfont}
\usepackage{textcomp}
\usepackage{multirow}
\usepackage{booktabs}
\usepackage{tikz}
\usepackage{overpic}
\usepackage{hyperref}
\usepackage{bm}
\usepackage{physics}
\usepackage{float}
\usepackage{xcolor}
\usepackage{caption}
\usepackage{subcaption}
\newcommand{\be}{\begin{equation}}
\newcommand{\ee}{\end{equation}}
\newcommand{\bea}{\begin{eqnarray}}
\newcommand{\eea}{\end{eqnarray}}
\newcommand{\ba}{\begin{align}}
\newcommand{\ea}{\end{align}}
\def\({\left(}
\def\){\right)}
\def\[{\left[}
\def\]{\right]}

\def\6{\partial}

\def\htt{\hat{\tau}}
\def\hn{\hat{n}}

\def\det{\textrm{det}}

\usepackage{calligra}
\DeclareMathAlphabet{\mathcalligra}{T1}{calligra}{m}{n}
\DeclareFontShape{T1}{calligra}{m}{n}{<->s*[2.2]callig15}{}

\subheader{\begin{flushright}
\texttt{IFT-UAM/CSIC-23-26}
\end{flushright}}

\title{Holographic entanglement as nonlocal magnetism}

\author[a]{Umut G\"ursoy,}
\author[b]{Juan F. Pedraza}
\author[a]{and Guim Planella Planas}
\affiliation[a]{Institute for Theoretical Physics, Utrecht University, 3584 CE Utrecht, The Netherlands}
\affiliation[b]{Instituto de F\'isica Te\'orica UAM/CSIC, Calle Nicol\'as Cabrera 13-15, Madrid 28049, Spain}
\emailAdd{u.gursoy@uu.nl}
\emailAdd{j.pedraza@csic.es}
\emailAdd{g.planellaiplanas1@uu.nl}

\abstract{The Ryu-Takayanagi prescription can be cast in terms of a set of microscopic threads that help visualize holographic entanglement in terms of distillation of EPR pairs. While this framework has been exploited for regions with a high degree of symmetry, we take the first steps towards understanding general entangling regions, focusing on AdS$_4$. Inspired by simple constructions achieved for the case of disks and the half-plane, we reformulate bit threads in terms of a magnetic-like field generated by a current flowing through the boundary of the entangling region. The construction is possible for these highly symmetric settings, leading us to a modified Biot-Savart law in curved space that fully characterizes the entanglement structure of the state. For general entangling regions, the prescription breaks down as the corresponding modular Hamiltonians become inherently nonlocal. We develop a formalism for general shape deformations and derive a flow equation that accounts for these effects as a systematic expansion. We solve this equation for a complete set of small deformations and show that the structure of the expansion explicitly codifies the expected nonlocalities. 
Our findings are consistent with numerical results existing in the literature, and shed light on the fundamental nature of quantum entanglement as a nonlocal phenomenon.}

\begin{document}
\maketitle
\flushbottom

%%%%%%%%%%%%%%%%%%%%%%%%%%%%%%%%%%%%%%
%%%%%%%%%%%%%%%%%%%%%%%%%%%%%%%%%%%%%%
\section{Introduction and summary}
%%%%%%%%%%%%%%%%%%%%%%%%%%%%%%%%%%%%%%
%%%%%%%%%%%%%%%%%%%%%%%%%%%%%%%%%%%%%%

Recent developments in quantum gravity have uncovered deep connections between spacetime and quantum information. The celebrated Ryu-Takayanagi (RT) formula \cite{Ryu:2006bv}, applicable in the context of holography, is arguably one of the most exciting results that sparked this type of research. This formula provides an elegant generalization of the Bekenstein-Hawking entropy-area relation for black holes \cite{Bekenstein:1973ur,Hawking:1974sw}, relating the area of particular codimension-2 surfaces in the bulk to the entanglement, or von Neumann entropy of subsystems in the dual CFT. A formal proof of the prescription was provided in \cite{Lewkowycz:2013nqa}. It has further been extended in a number of ways, including a fully covariant version \cite{Hubeny:2007xt,Dong:2016hjy}, a generalization to gravity theories beyond Einstein \cite{Dong:2013qoa,Camps:2013zua}, and a generalization that includes bulk quantum corrections \cite{Faulkner:2013ana,Engelhardt:2014gca}.

Explorations of the RT prescription have revealed that spacetime itself may be an emergent concept, giving rise to the slogan `entanglement=geometry' \cite{VanRaamsdonk:2009ar,VanRaamsdonk:2010pw,Bianchi:2012ev,Maldacena:2013xja,Balasubramanian:2014sra}. Indeed,
one of the most interesting applications of this prescription is the program of metric reconstruction, which have been studied in a variety of settings 
\cite{Czech:2012bh,Balasubramanian:2013lsa,Myers:2014jia,Hubeny:2014qwa,Czech:2014wka,Headrick:2014eia,Czech:2014ppa,Czech:2015qta,Faulkner:2018faa,Roy:2018ehv,Espindola:2017jil,Espindola:2018ozt,Balasubramanian:2018uus,Bao:2019bib,Jokela:2020auu,Bao:2020abm}. More generally, it has been understood that the dynamics of the bulk metric, i.e., the Einstein equations, map to the laws governing the entanglement entropy under changes in the CFT state or the CFT Hamiltonian \cite{Lashkari:2013koa,Faulkner:2013ica,Swingle:2014uza,Caceres:2016xjz,Faulkner:2017tkh,Haehl:2017sot,Rosso:2020zkk,Taylor:2021jqi}. As a result, not only geometry but gravity itself appear to be emergent in holographic settings.  Many other connections between gravity and quantum information have been inspired by the RT formula, including tensor networks \cite{Swingle:2009bg,Hayden:2016cfa,Bao:2018pvs,Jahn:2021uqr}, quantum error correction \cite{Almheiri:2014lwa,Pastawski:2015qua,Dong:2016eik,Harlow:2016vwg}, quantum computation \cite{Susskind:2014rva,Brown:2015bva,Brown:2015lvg,Caputa:2017yrh,Couch:2016exn} and quantum teleportation \cite{Gao:2016bin,Maldacena:2017axo,Brown:2019hmk,Freivogel:2019lej,Freivogel:2019whb}.

Recent work by Freedman and Headrick demonstrated that the RT formula is amenable to a dual description in terms of flows, divergenceless norm-bounded vector fields, or alternatively, a collection of Planck-thickness ``bit threads'' \cite{Freedman:2016zud}. This novel prescription has revealed hitherto undiscovered properties of holographic entanglement and has given them a neat information-theoretic interpretation in terms of distillation of EPR-type entanglement  \cite{Headrick:2017ucz,Chen:2018ywy,Cui:2018dyq,Hubeny:2018bri,Agon:2018lwq,Ghodrati:2019hnn,Du:2019emy,Bao:2019wcf,Harper:2019lff,Agon:2019qgh,Du:2019vwh,Agon:2020mvu,Headrick:2020gyq,Lin:2020yzf,Lin:2021hqs,Lin:2022aqf,Mintchev:2022fcp,Lin:2022agc,Lin:2022flo}.
Like the RT formula, bit threads have been extended to covariant settings \cite{Headrick:2022nbe}, non-Einstein gravity theories \cite{Harper:2018sdd}, and when bulk quantum corrections are considered \cite{Agon:2021tia,Rolph:2021hgz}. However, with the exception of \cite{Agon:2018lwq,Agon:2020mvu}, not a lot of work has been done in finding bit thread solutions for specific cases of interest.  This is a little surprising, as bit threads define a convex program that should be amenable to numerical computation, more so than the standard RT prescription. Indeed, depending on the shape, finding RT surfaces can be computationally very demanding \cite{Fonda:2014cca,Fonda:2015nma,Seminara:2018pmr,Cavini:2019wyb}, hence the need for the development of new tools.

In this paper we take steps toward filling this gap. We reformulate the vector field that describes bit threads in terms of a bulk magnetic-like field emanating from inside the entangling region and terminating on the outside of this region. The vector field $V$ that describes bit threads in the classical limit must be divergenceless, $\nabla\cdot V=0$, and norm bounded, $|V|\leq 1$, with the maximum attained on the RT surface.  Magnetic fields provide a convenient description in this context as they are automatically divergenceless in the absence of magnetic sources. Indeed, they can be thought of as generated only by a hypothetical electric current flowing through the boundary of the entangling surface via a modified Biot-Savart law, see Fig.~\ref{Vhalfplane}. We show that the norm of these modified magnetic fields attains a {\em local} maximum on the RT surface. This local maximum is also the {\em global} one in maximally symmetric cases, such as when the entangling surface is a half-plane or a disk. Indeed, we 
demonstrate that the magnetic field lines match precisely the bit threads obtained from a prescription based on geodesics \cite{Agon:2018lwq,Agon:2020mvu} in such maximally symmetric cases. 

However, we also show that magnetic fields do not correspond to a global maximum in the general case, i.e., for entangling regions with arbitrary shapes. In the second part of our paper we, therefore, move on to provide another calculation scheme to determine the RT surface and the corresponding bit threads, which continue to hold for arbitrary shape deformations. Inspired by the geodesic construction, we first derive equations of motion for the bulk metric in the Gaussian normal coordinates adapted to the RT surface and then obtain the flow equations that characterize arbitrary shape deformations in a perturbative fashion. Quite interestingly, for conformal field theories, this perturbative scheme directly mapped to a Witten-diagrammatic prescription whereby the metric on the deformed RT surface is determined by propagating the deformation of the entangling region on the boundary to the bulk point via bulk-to-boundary and bulk-to-bulk AdS propagators convoluted through bulk vertices that are obtained from the aforementioned flow equations, see Fig.~\ref{Wittendiagrams} for an illustration. We then present a general formula for the vector field that corresponds to the bit threads based on Gaussian construction. This method establishes --- at least when the bulk is AdS and the corresponding CFT state is the vacuum state --- that, space-like geodesics generally provide a valid bit thread configuration. It also confirms their representation as magnetic field lines which we present in the first part of our paper, as the lowest-order contribution in the aforementioned perturbative series. The higher-order terms in this series then provide nonlocal corrections to their representation as magnetism.   

The rest of the paper is organized as follows. In section \ref{sec:magnetic}, we present our description of bit threads as magnetic field lines and determine the limitations of this description. In section \ref{sec:deform}, we present our more general perturbative scheme to compute the RT surfaces and corresponding bit threads for entangling regions of arbitrary shape. In section \ref{sec:conclusion}, we conclude with a discussion of our findings and an outlook on their possible extensions. We relegate some computational details to the two appendices. 

%%%%%%%%%%%%%%%%%%%%%%%%%%%%%%%%%%%%%%
%%%%%%%%%%%%%%%%%%%%%%%%%%%%%%%%%%%%%%
\section{Bit threads as magnetic flows\label{sec:magnetic}}
%%%%%%%%%%%%%%%%%%%%%%%%%%%%%%%%%%%%%%
%%%%%%%%%%%%%%%%%%%%%%%%%%%%%%%%%%%%%%

Let us begin by reviewing the proposal of \cite{Freedman:2016zud}.  The idea is to recast the RT prescription using flows, divergenceless norm-bounded vector fields, or alternatively, a collection of Planck-thickness bit threads. More specifically, according to this proposal, we can compute entanglement entropy in holographic settings as a maximizing flux, such that
\be\label{BTpres}
S[\mathcal{A}]=\frac{1}{4G_N}\, \max_{V\in \mathcal{F}}\int_{\mathcal{A}} V\,,\qquad {\cal F}\equiv\{V\, \vert\, \nabla\cdot V=0,\, |V|\leq 1\}\,.
\ee
The equivalence with the RT prescription follows from the continuum version of the max-flow min-cut theorem of network theory. More generally, one can arrive at such a prescription using convex optimization and strong duality of linear programs \cite{Headrick:2017ucz}.

Discretized flows are interpreted as bit threads, microscopic 
fibers that codify the entanglement between pairs of degrees of freedom in the dual CFT state. The arrangement of threads, however, is highly non-unique. Intuitively, such degeneracy can be associated with choices of microstates that give rise to the same amount of entanglement between the region of interest and its complement. So, different bit thread solutions codify different configurations built out from EPR-type entanglement. A handful of methods for computing particular thread configurations were introduced in \cite{Agon:2018lwq,Agon:2020mvu}. Among these, the so-called ``geodesic bit threads'' were recently interpreted in terms of modular conjugations in the dual conformal field theory \cite{Mintchev:2022fcp}. In the following, we will consider simple examples of such bit thread configurations. Later in the same section, we will provide a novel bulk interpretation in terms of magnetic-like fields, however, we will argue that such reinterpretation will generally require corrections when considering entangling regions with less symmetry.

\subsection{Simple analytic constructions in AdS$_4$\label{sec:geoBTs}}
A simple method to construct flows is to take space-like geodesics as integral curves and then use Gauss's law to fix the norm. This works for disks and strips in pure AdS$_4$, or more general asymptotically AdS$_4$ geometries as long as the bulk satisfies certain energy conditions \cite{Agon:2018lwq}.
Following the same method, we will start by constructing a flow for the case of the semi-infinite plane in empty AdS. For the sake of comparison, we will also show the result for the case of the disk, which was derived analytically in \cite{Agon:2018lwq}. 

Before starting, we recall that in general CFTs, the modular hamiltonian of the semi-infinite plane is related to that of the sphere by a conformal transformation. Therefore, we expect to be able to derive an analytic flow for this case as well.

To begin with, consider the minimal surface associated with a semi-infinite plane $A$, defined as $x_1\equiv x\in(-\infty,0]$, in pure AdS$_4$. In Poincar\'e coordinates, constant-$t$ slices of the metric have the following line element
\bea
ds^2=\frac{1}{z^2}\(d\vec{x}^2+dz^2\)\,.
\eea
Given the translation invariance along $x_2$, we can suppress this coordinate so that the problem can be effectively reduced to two dimensions. The minimal surface $m(A)$, parametrized by $(x_m,z_m)$, is simply a vertical surface with
\bea\label{minsphere}
x_m=0\,.
\eea
The outward-pointing unit normal vector $\hn_m$ at a point $(x_m,z_m)$, is given by
\bea\label{nasphere}
\hn_m^{a}=\(z_m,0\)\,,
\eea
where the index $a$ here runs over the coordinates $(x,z)$.

\begin{figure}[t!]
\centering
 \includegraphics[width=2.84in
 ]{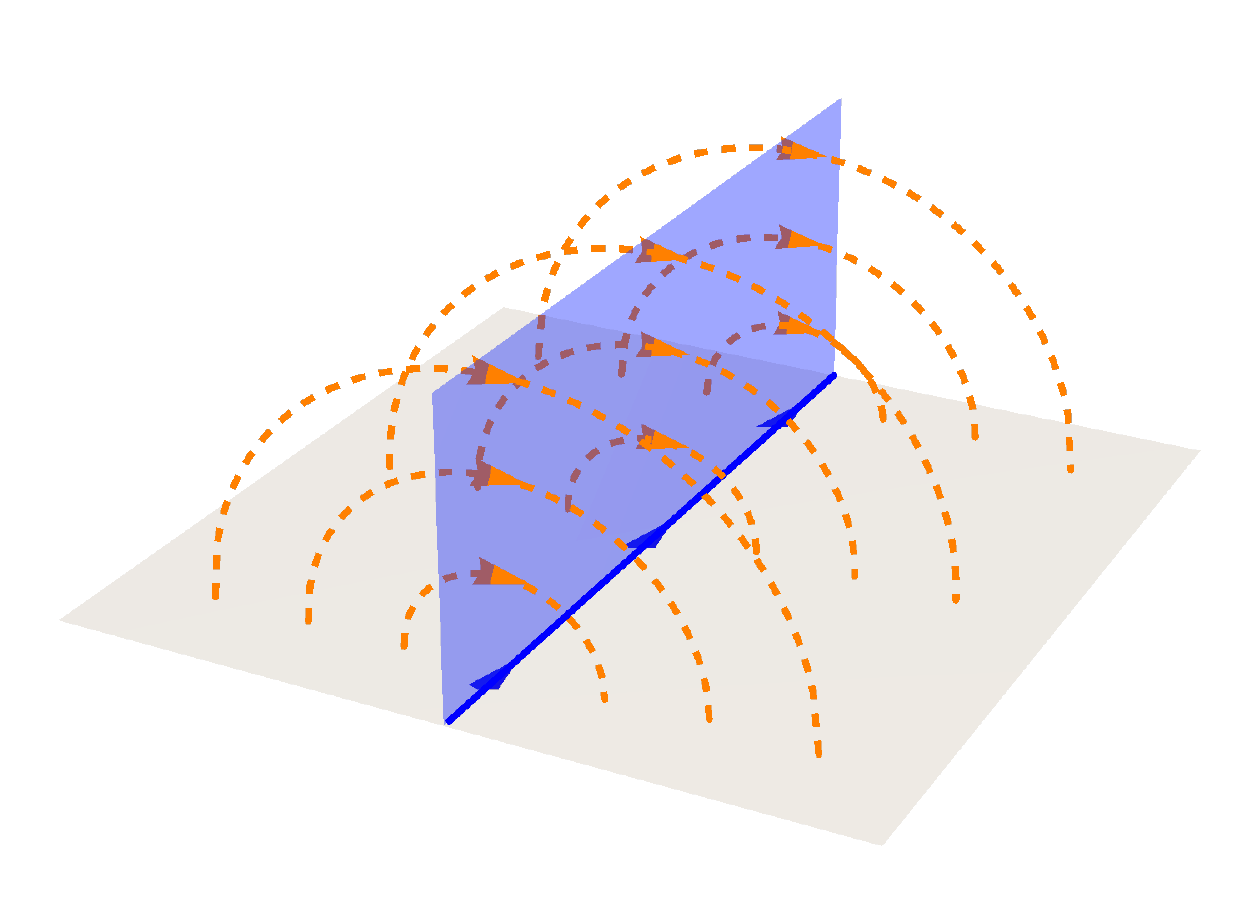}
  \includegraphics[width=3in
 ]{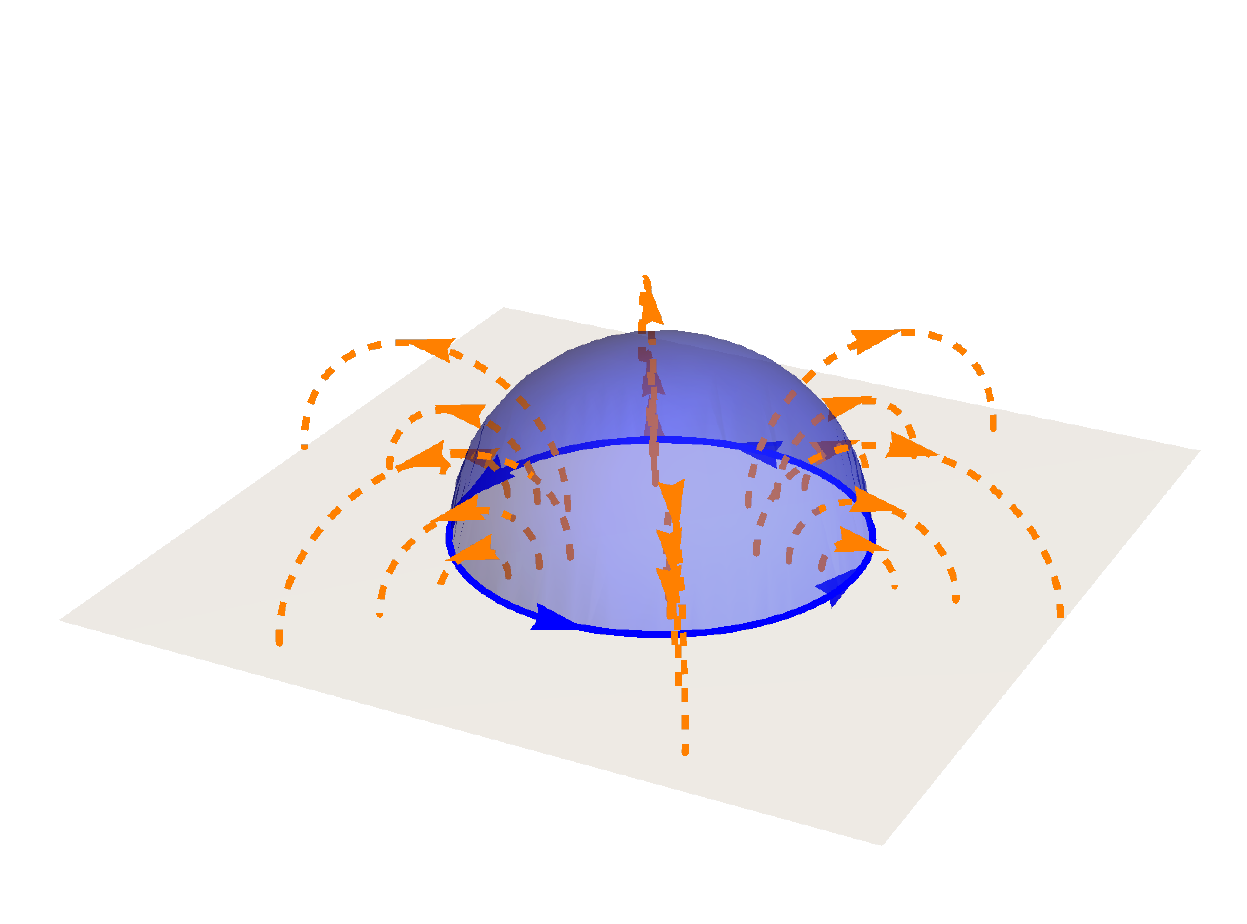}
 \begin{picture}(0,0)
\put(-365,25){{\small $x_1$}}
\put(-258,38){{\small $x_2$}}
\put(-148,25){{\small $x_1$}}
\put(-37,38){{\small $x_2$}}
\put(-408,77){{\small $z$}}
\put(-193,77){{\small $z$}}
\put(-320,58){{\small $I$}}
\put(-88,44){{\small $I$}}
\put(-405,80){{\LARGE $\uparrow$}}
\put(-190,80){{\LARGE $\uparrow$}}
\end{picture}
 \caption{Threads or integral lines for the vector field corresponding to a semi-infinite plane and a disk, respectively. In both cases, the edge of the entangling region is drawn in solid blue, at the boundary of AdS. The minimal surfaces extend into the bulk and are also shown in (lighter) blue. The flow lines suggest a magnetic-like interpretation, which may be generated by an electric current $I$ flowing along the edge of the entangling region.}\label{Vhalfplane}
\end{figure}
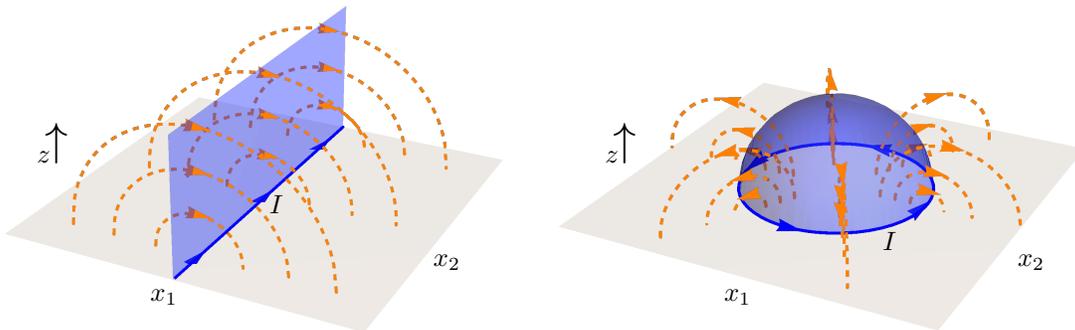

Consider now the space of geodesics that lie on the $(x,z)$ plane that intersects $m(A)$ at $(x_m,z_m)$ with tangent vector $\hat{\tau}$ equal to the normal $\hn_m$ at that point. It is easy to check that the set of relevant geodesics is given by the one-parameter family of circumferences defined implicitly by
\bea\label{geodesic}
x^2+z^2=R_s^2\,,
\eea
i.e., semi-circles centered at $x=0$ with radii $R_s$ (see Fig.~\ref{Vhalfplane} below for an illustration). The tangent vector with unit norm at an arbitrary point along one of these geodesics is given by
\bea\label{tau}
\hat{\tau}^a=\frac{z}{R_s}\(z , -x\)\,.
\eea
Enforcing that $\hat{\tau}=\hn_m$ at a point $(r_m,z_m)$ on the minimal surface leads to the simple condition
\bea\label{80}
R_s=z_m\,.
\eea
Plugging (\ref{80}) into  (\ref{geodesic}) and (\ref{tau}) we obtain an implicit expression for the family of geodesics orthogonal to $m(A)$ with the correct parametrization,
\be\label{geo2}
x^2+z^2=z_m^2\,,
\ee
and tangent vector
\bea\label{tau2}
\hat{\tau}^a=\frac{z}{z_m}\(z , -x\)\,.
\eea
It is straightforward to check that these geodesics are nested and non-intersecting, which validates our choice as integral curves \cite{Agon:2018lwq}.

We can now proceed to find the appropriate norm of the vector field $|V|$. First, we compute the orthogonal metric at different points along the minimal surface,
\bea\label{orthospheres}
h_{ab}(z_m;x,z)=g_{ab}-\hat{\tau}_a \hat{\tau}_b\,,
\eea
where $\hat{\tau}$ is the unit tangent vector, given in (\ref{tau2}). A brief calculation leads to
\bea\label{dsperp1}
ds_\perp^2\equiv h_{ab}dx^a dx^b=\frac{1}{z^2z_m^2}\left(xdx+z dz \right)^2\,.
\eea
Now, we can use the geodesic equation (\ref{geo2}) to $(i)$ eliminate one of the variables, and $(ii)$ express (\ref{dsperp1}) as a differential along the transverse coordinate, namely $dz_m$. By doing so, and restoring the coordinate $x_2$, we find that
\bea
ds_\perp^2=\frac{1}{z^2}\(dz_m^2+dx_2^2\) \,.
\eea
The magnitude of the vector field reads
\bea\label{magVsph}
|V|=\frac{\sqrt{h(z_m;x_m,z_m)}}{\sqrt{h(z_m;x,z)}}=\(\frac{z}{z_m}\)^{2}\,.
\eea
Finally, it would be convenient to express our vector field as a function of $(x,z)$ without reference to the minimal surface. This can be achieved by solving for $z_m=z_m(x,z)$ from the geodesic (\ref{geo2}) and plugging it back into the relevant equations. A short calculation leads to
\be
|V|=\frac{z^2}{x^2+z^2}\,,
\ee
and
\bea\label{tau3}
\hat{\tau}^a=\frac{z}{\sqrt{x^2+z^2}}\(z  ,-x\)\,.
\eea
Putting it all together, we find that the full vector field $V=|V|\htt$ is given by
\bea\label{Vd3}
V^a=\(\frac{z}{\sqrt{x^2+z^2}}\)^{3}\(z , -x\)\,.
\eea
Notice that, in the above expressions, we have suppressed the component of the vector in the transverse direction $x_2$, which is identically zero (by translation invariance).

For comparison's sake, here we show the resulting vector field for the case of the disk, which was worked out in \cite{Agon:2018lwq}:
\bea\label{Vd2}
V^a=\(\frac{2Rz}{\sqrt{(R^2+r^2+z^2)^2-4R^2r^2}}\)^{3}\(\frac{r z}{R}\, , \frac{R^2-r^2+z^2}{2R} \)\,.
\eea
with
\be
|V|=\(\frac{2Rz}{\sqrt{(R^2+r^2+z^2)^2-4R^2r^2}}\)^{2}\,,
\ee
\be
\htt^a \equiv\frac{V^a}{|V|}=\frac{2Rz}{\sqrt{(R^2+r^2+z^2)^2-4R^2r^2}}\( \frac{r z}{R}\, ,\, \frac{R^2-r^2+z^2}{2R}\)\,.
\ee

In Figure \ref{Vhalfplane} we show plots of some of the integral curves for both of the vector fields discussed above. Quite remarkably, they are reminiscent of the magnetic flow lines generated by an electric current flowing along the edge of the entangling region. In the remainder of this section, we will explore this proposal more concretely.

\subsection{Magnetic interpretation}

The preceding examples point to a natural conjecture that we would like to put forward. In particular, we would like to understand if a bit thread configuration constructed from geodesics can be interpreted as a magnetic field sourced by a current, and if so, how universal would this proposal be. To tackle this question, we need to solve a boundary value problem in curved space and find a vector field with maximum flux through the given region while being norm bounded from above everywhere in the bulk. 

This problem is highly non-trivial when looking for global maxima. However, we can write an action together with the corresponding Lagrange multipliers and find local extrema of the flux by solving the corresponding Euler-Lagrange equations. This procedure naturally leads us to an equation reminiscent of Ampere's law, suggesting a magnetic-like field may be a possible solution. Indeed, we will show that a modified Biot-Savart law is in fact a valid solution for the bit threads and thus a local extremum of the flux. We conclude the section by commenting on the possible universality of the prescription or the lack thereof.

\subsubsection{Exact Maxwell's equation on the RT surface}

We need to find a divergenceless vector field that maximizes the flux through the region of interest. Let us start by introducing a $1$-form $A$ on a constant-$t$ slice of Poincaré-AdS space,
\begin{equation}
 	A= A_x(x,y,z) \dd x + A_y(x,y,z) \dd y + A_z(x,y,z)\dd z\,.
 \end{equation}
The most general divergenceless vector field can be written as
\begin{equation}
	V=\star\;\dd A\,,
\end{equation}
where $\star$ stands for the Hodge star operator which maps $p$-forms to $(d-p)$-forms,
\begin{equation}
	(\star\; \alpha)_{i_{p+1}\dots i_{d}}= \frac{\sqrt {|\det g|}}{p!} g^{i_1 j_1}\dots g^{i_p j_p }\alpha_{j_1 \dots j_ p} \varepsilon_{i_1\dots i_n}\,.
\end{equation}
Its flux through a surface $\mathcal{A}$ can be written as
\begin{equation}
 	\int_\mathcal{A} \star\; V = \int_ \mathcal{A} \dd A = \int_ {\partial\mathcal{ A}} A = \int \dd s  \left( A_x \frac{\dd x(s)}{\dd s} + A_y \frac{\dd y(s) }{\dd s}   \right)\,,
 \end{equation}
where the functions $x(s)$ and $y(s)$ parametrize the curve $\partial \mathcal{A}$. We have used Stokes' theorem for integration on a manifold and taken $\mathcal{A}$ to be located at some $z=\epsilon\to0$. Note here that  $\star \star \dd A= \dd A$ since $A$ is a one-form in a $3$ dimensional space with positive definite metric.

To maximize this flux we introduce a Lagrange multiplier $\phi$ enforcing the constraint. This leads us to the following action,
 \begin{equation}
 		S= \int_{\partial \mathcal{A}} A -	\int_ {\mathcal{M}} \phi \left( 1 - \psi^2 - \left|\star \; \dd A\right|_{\text{AdS}}^2 \right) \dd v\,,
 \end{equation}
 with the norm $|\cdot|_{\text{AdS}}$ computed using the AdS metric and the volume element $dv$ including the metric determinant in AdS. Here, we have introduced an auxiliary slack variable $\psi$ which allows for $\left|\star \; \dd A\right|_{\text{AdS}}^2 < 1$, i.e., it accounts for the fact that the constraint is an inequality, rather than an equality \cite{BookConvex}.  The modulus of $V$ is easily computed to be
\be
 	\left|\star \; \dd A\right|_{\text{AdS}}^2=z^4\left| \nabla \times \boldsymbol{A}\right|_{\mathbb{R}^3}^2\,,
\ee
where $\boldsymbol{A}=(A_x,A_y,A_z)$ is a vector in $\mathbb{R}^3$ and both the curl and the norm are computed as in a normal Cartesian flat space. From here on, unless specified otherwise, $|\cdot|$ will always denote the norm in flat space. In terms of this vector field, our action becomes,
\begin{align}
S&= \int \dd s \;\boldsymbol{A} \cdot \frac{\dd \boldsymbol{r }}{\dd s} - \int \frac{\dd ^3 r}{z^3}  \; \phi\left( 1- \psi^2 - z^4 \left|\nabla \times \boldsymbol{A}\right|^2  \right) \nonumber\\
&=\int \dd^3 r \left( \boldsymbol{A}\cdot\int \dd s \;\delta^{(3)}\left(\boldsymbol{r}- \boldsymbol{r}(s)\right)  \frac{\dd \boldsymbol{r }}{\dd s} - \frac{1}{z^3}\phi \left( 1- \psi^2 - z^4 \left|\nabla \times \boldsymbol{A}\right|^2 \right)   \right)\,,
\end{align}
which must be varied over $\boldsymbol{A}$, $\phi$ and $\psi$. The corresponding Euler-Lagrange equations for $\phi$ and $\psi$ are, respectively,
\begin{equation}
\label{eq:constraints}
	\frac{1}{z^3}\left( 1- \psi^2 - z^4 \left|\boldsymbol{B}\right|^2  \right) =0\,, \qquad\qquad \frac{2}{z^3 } \phi \psi=0\,,
\end{equation}
where $\boldsymbol{B}\equiv \nabla \times \boldsymbol{A}$. For $\boldsymbol{A}$ we get
\begin{equation}
	\frac{\partial \mathcal{L}}{\partial A_i}= \int\dd s \; \delta^{(3)} \left(\boldsymbol{r}- \boldsymbol{r}(s)\right) \frac{\dd r_ i}{\dd s }\,,  \qquad\qquad  \partial_ i\frac{\partial \mathcal{L }}{\partial (\partial_ i A_ j)} =-\left( \nabla \times   \left( z \phi\, \boldsymbol{B} \right)  \right)_j\,,
\end{equation}
which gives rise to the following Euler-Lagrange equation
\begin{equation}
\label{eq:modifiedMaxwell}
	\nabla\times \left( z \phi\,\boldsymbol{B} \right) = - \int \dd s \; \delta^{(3)} \left( \boldsymbol{r}- \boldsymbol{r}(s) \right) \frac{\dd \boldsymbol{r}}{\dd s }\equiv \boldsymbol{J}\,.
\end{equation}
Here the curl is again taken in Cartesian flat space.

The two equations \eqref{eq:constraints} amount to requiring
 \begin{equation}
 	z^4\,\left|\boldsymbol{B}\right|^2\le 1\,, \qquad \qquad \phi= \frac{1}{z} f(x,y, z)\chi _m\,,
 \end{equation}
 where $\chi_ m$ is the indicator function of the RT surface $m$, which is $1$ when a point belongs to the RT surface
 and $0$ otherwise. Clearly, the inequality will be saturated on the RT surface. This follows from the fact that $\psi$ should vanish whenever $\phi$ is nonzero and the requirement that $\psi\in \mathbb{R}$. The factor of $z^{-1}$ on $\phi$ is just a matter of the definition of $f$ and has been added to simplify equation \eqref{eq:modifiedMaxwell}. Using these relations on \eqref{eq:modifiedMaxwell} we get
 \begin{equation}
 \label{eq:modifiedMaxwellRT}
 	\chi_m \nabla\times  \left(f \boldsymbol{B}  \right)+\nabla \chi_m \times (f\boldsymbol{B})= \chi_m \nabla\times  \left(f \boldsymbol{B}  \right) =\boldsymbol{J}\,.
 \end{equation}
Here we have used that $\nabla \chi_m$ is parallel to $\boldsymbol{B}$, as both must be perpendicular to the RT surface. Thus, we get a modified version of Maxwell's equations for a magnetostatic situation but, strictly speaking, they only need to be fulfilled on the RT surface, not on the whole space. Assuming $\boldsymbol{B}$ could be extended everywhere in the bulk, then, one would recover the components for the bit thread vector field in Poincar\'e-AdS as
\begin{equation}\label{eq:transAdS}
    V^i=z^2 \delta^{ij}(\star d A)_j=z^3 \delta^{ij} (\nabla\times \boldsymbol{A})_j= z^3 \boldsymbol{B}_i\,.
\end{equation}

The above result is intuitive, as bit threads are generally floppy on $\mathcal{M}$ and can generally be deformed, except in the vicinity of the RT surface where they become rigid. This is due to the fact that threads are maximally packed on $m$, which disfavors any deformation. In the next subsection, we will try to extend (\ref{eq:modifiedMaxwellRT}) to the full manifold, by suitably picking a propagator for the magnetic field. Such extrapolation may not be possible in general, as entanglement is inherently nonlocal, while this prescription assumes locality. However, we will show that for some highly symmetric cases this prescription works, leading to a complete characterization of the entanglement pattern of the quantum state.

\subsubsection{Generalized Biot-Savart law from flux maximization and boundedness}

Given a specific current on $\partial \mathcal{M}$, the next step is to propose a suitable recipe for computing  a magnetic-like field that extends over $\mathcal{M}$ while satisfying (\ref{eq:modifiedMaxwellRT}) on the RT surface. In standard electromagnetism, the Biot-Savart law states that the magnetic field obeys the standard inverse-square law
\begin{equation}\label{BSLawforB}
   \mathbf{ B(r})=\frac{\mu_0}{4 \pi}\int \frac{\mathbf{J}(\mathbf{r'})\times(\mathbf{\hat{r}}-\mathbf{\hat{r}'})}{|\mathbf{r}-\mathbf{r'}|^2} \dd^3 r',
\end{equation}
and satisfies the divergenceless condition $\nabla\cdot \mathbf{ B(r})=0$. To make this point evident, we can  recast (\ref{BSLawforB})
as $\mathbf{B(r})=\nabla\times\mathbf{ A(r)}$, where
\begin{equation}\label{BSLawforA}
 \mathbf{ A(r})\equiv\frac{\mu_0}{4 \pi}\int \frac{\mathbf{J}(\mathbf{r'})}{|\mathbf{r}-\mathbf{r'}|} \dd^3 r'\,,
\end{equation}
which corresponds to a particular gauge. In terms of the vector potential $\mathbf{ A(r)}$ the divergenceless condition is manifest, which is particularly convenient and easier to generalize. This expression for the magnetic field will obviously be a solution to our generalized equation with a constant $f$. Moreover, not only does it solve the equation on the RT surface but it also solves it everywhere in the bulk. 

However, we run into problems when we examine the norm bound more closely. To see this, consider the behavior of $|\boldsymbol{B}|$ in the vicinity of the curve $\partial \mathcal{A}$ lying on the boundary,
\begin{equation}
   \lim_{r\to r(0)} \frac{z^2}{\ell^2} \left|\boldsymbol{B}\right|=\lim_{\epsilon\to 0}\left|\int \dd s \frac{\epsilon^2}{\ell^2} \frac{(r(0)-r(s)+ \epsilon \hat{z})\times \frac{\dd r(s)}{\dd s} }{\left|r(0)-r(s)+ \epsilon \hat{z}\right|^3}\right|\,.
\end{equation}
When $\epsilon$ is small enough, the only contribution to this integral comes from the values of $s$ close to $0$. In this region we can approximate $r(s)$ by the leading term in a Taylor series and extending the range of integration to infinity will not change the result since once again only small values of $s$ contribute. Overall one finds,
\begin{equation}
    \lim_{r\to r(0)}z^2\left|\boldsymbol{B}\right| = \lim_{\epsilon\to 0} \left|\int \dd s \epsilon^2\frac{ \epsilon \hat{z}\times \left. \frac{\dd r}{\dd s }\right|_{0} }{\left( \left| \left. \frac{\dd r}{\dd s}\right|_{0} \right|^2 s^2 + \epsilon^2 \right)^ {\frac{3}{2}}}\right|= \lim_{\epsilon\to 0}    2 \epsilon=0\,.
\end{equation}
While this certainly respects the norm bound, $r(s)$ is part of the RT surface, and thus, the magnetic field there should saturate the bound for any bit thread configuration maximizing the flux. As a consequence, the usual magnetic field given by (\ref{BSLawforB}) cannot maximize the flux. The simplest generalization which solves this problem is to consider a modified Biot-Savart law of the form
\begin{equation}
    \boldsymbol{B} = C\int  \frac{\boldsymbol{J}(r')\times(\boldsymbol{r}-\boldsymbol{r}')}{\left|\boldsymbol{r}- \boldsymbol{r}' \right|^n} \dd^3 r' \,,
\end{equation}
coming from a vector potential where we have modified the power of $\left|\boldsymbol{r}- \boldsymbol{r}' \right|$ appearing in the integral and where $C$ and $n$ are constants to be determined. The value of $n$ can be fixed by imposing covariance with respect to conformal transformations. Obviously, this expression will be covariant under rotations and translations, however, we also want the magnetic field to transform properly under dilations and special conformal transformations. For dilations, this condition reduces to
\begin{align}
    V_{\boldsymbol{r_0}}^i(x^i)\partial_{x^i}&=V_{\lambda \boldsymbol{r_0}}^i(\lambda x^i)\partial_{\lambda x^i} = C (\lambda z)^3 \int \frac{(\boldsymbol{J}_{\lambda \boldsymbol{r_0}}(r')\times(\lambda \boldsymbol{r}-\boldsymbol{r}'))^i}{\left|\lambda \boldsymbol{r}- \boldsymbol{r}' \right|^n} \dd^3 r'\partial_{\lambda x^i}= \lambda^{4-n} V_{\boldsymbol{r_0}}^i(x^i)\partial_{x^i}\,,
\end{align}
where the subindices indicate the curve we are using to compute the vector field. Thus, we find that covariance with respect to dilations implies $n=4$. As for special conformal transformations, one can show that they will also respect covariance with $n=4$ though the calculation is much more lengthy. Fixing the coefficient requires looking again at the behavior of the modulus of the magnetic field when going near the sources. Repeating the same analysis as before yields
\begin{equation}
    1=\lim_{r\to r(0)}z^2 \left|\boldsymbol{B}\right| =  \left|C\right|\lim_{\epsilon\to 0}  \frac{\sqrt{\pi} \Gamma \left( \frac{n-1}{2} \right) }{\Gamma \left( \frac{n}{2} \right) } \epsilon^{4-n}\qquad \Rightarrow\quad  n=4\,,\quad \left|C\right|= \frac{2}{\pi}\,,
\end{equation}
We see that apart from fixing the coefficient the near-source behavior also requires $n=4$. Then, we obtain the prescription
\begin{equation}
   \label{eq:magneticfield}
    \boldsymbol{B} = \frac{2}{\pi}\int   \frac{\boldsymbol{J}(r')\times(\boldsymbol{r}- \boldsymbol{r}')}{\left|\boldsymbol{r}- \boldsymbol{r}' \right|^4}\dd^3 r' \,,
\end{equation}
with the corresponding vector potential
\begin{equation}
\label{ModifiesBSlawA}
    \boldsymbol{A}= \frac{1}{\pi}\int \frac{\boldsymbol{J}(\boldsymbol{r}')}{\left|\boldsymbol{r}- \boldsymbol{r}'\right|^2}\dd^3 r' \,.
\end{equation}
With this generalization of $\boldsymbol{B}$, it is no longer obvious that there is a choice of $f$ such that equation \eqref{eq:modifiedMaxwellRT} is satisfied. However, a straightforward analysis shows that one may fix $f(x,y,z)= \frac{z}{2}$ and the equation is satisfied everywhere. For details see appendix \ref{Fluxmaximization}.

\subsubsection{Explicit results and match with geodesic bit threads}

Let us now use our modified Biot-Savart law \eqref{ModifiesBSlawA}.
We will analyze two cases, the straight current, and the circular current. In both cases, we will compute $\mathbf{ B}=\nabla\times\mathbf{ A}$ and then transform the field into AdS space. Finally, we will compare our results with those obtained using geodesic bit threads.

\subsubsection*{The straight current}
We will start with a straight current flowing along the $y$-axis, which is the desired configuration when the entangling region corresponds to a half-plane ($x<0$). In this case, we have
\be
\mathbf{J}(\mathbf{r'})=  \delta(x')\delta(z')\hat{\boldsymbol{y}}\,.
\ee
A brief calculation shows that
\begin{align}
    \mathbf{A(r)}=  \hat{\boldsymbol{y}} \lim_{R\to\infty} \int_{-R}^{R} \frac{\dd y'}{x^2 + (y-y')^2 + z^2}= \frac{1}{\sqrt{x^2+z^2}} \hat{\boldsymbol{y}}\,,
\end{align}
so that
\be
\mathbf{B(r)}=\frac{1}{(x^2+z^2)^{3/2}}\left(z \hat{\boldsymbol{x}}-x \hat{\boldsymbol{z}}\right)\,.
\ee
Translating this result into AdS, using (\ref{eq:transAdS}), we obtain
\begin{equation}
    V= \left( \frac{z}{\sqrt{x^2+z^2}} \right)^3 \left( z \partial_x -x \partial_z \right)
\end{equation}
which exactly matches the result (\ref{Vd3}) obtained with geodesic bit threads.

\subsubsection*{The circular current}
The second example is the circular current, which is the expected configuration when the entangling region corresponds to a disk. In this case, it is convenient to work in cylindrical coordinates to represent the source, $x'=r'\cos(\phi)$ and $y'=r'\sin(\phi)$, so that
\be
\mathbf{J}(\mathbf{r'})= \delta(z')\delta(r'-R)\hat{\boldsymbol{\phi}}\,,
\ee
where $\hat{\boldsymbol{\phi}}=-\sin(\phi)\hat{\boldsymbol{x}}+\cos(\phi)\hat{\boldsymbol{y}}$.
After some algebra, we arrive at
\be
    \mathbf{A(r)}= \frac{R}{\pi} \int_{0}^{2\pi} \dd\phi\frac{-\sin(\phi)\hat{\boldsymbol{x}}+\cos(\phi)\hat{\boldsymbol{y}}}{(x-R\cos(\phi))^2+(y-R\sin(\phi))^2+z^2}\,.
\ee
which gives the following magnetic-like field
\be
    \mathbf{B(r)}=\frac{2 R}{\pi} \int_{0}^{2\pi} \dd\phi \frac{z \cos(\phi)\hat{\boldsymbol{x}}+z\sin(\phi)\hat{\boldsymbol{y}}+(R - x \cos(\phi) - y \sin(\phi))\hat{\boldsymbol{z}}}{[(x-R\cos(\phi))^2+(y-R\sin(\phi))^2+z^2]^2}\,.
\ee
Exploiting the rotational symmetry allows us to compute the above integrals. Translating the result into AdS, using (\ref{eq:transAdS}), one finds
\begin{equation}
    V= \left( \frac{2 R z}{\sqrt{(R^2+r^2+z^2)-4 R^2 r^2}} \right)^3 \left( \frac{r z}{R} \partial_r + \frac{R^2-r^2+z^2}{2R} \partial_z \right).
\end{equation}
Thus perfectly matches the result (\ref{Vd2}) obtained with geodesic bit threads.

\subsection{A comment on (non-)universality}
\label{sec:non_universality}
Remarkably, the modified Biot-Savart law,
\be\label{biots-law}
\mathbf{A(\mathbf{r})}= \frac{1}{\pi}\int \frac{\mathbf{J(r') }\dd^3 r' }{|\mathbf{r-r'}|^2}\,, 
\ee
with  $\mathbf{B(\mathbf{r})}=\nabla \times \mathbf{A(\mathbf{r})}$, works for both, the straight current (semi-infinite plane) and the circular current (disk), giving a flow that matches the one constructed from space-like geodesics. This suggests that a possible universal behavior may exist for arbitrarily shaped regions, which we will now investigate.

The simplest consistency check is to compute the magnetic field that follows from the law (\ref{biots-law}) for an entangling region with slightly less symmetry. For concreteness, we have considered the ellipse as a study case, defined as
\be
(x,y) \in \mathbb{R} \left| \left(\frac{x}{a}\right)^2 +\left(\frac{y}{b}\right)^2 = 1\right.\,.
\ee
In this case, we have
\be
\mathbf{J(r)}= \int\delta(z) \delta(x-a \cos (\phi)) \delta(y-b\sin (\phi))(-a \sin (\phi)\hat{\boldsymbol{x}}+ b \cos (\phi)\hat{\boldsymbol{y}}) d \theta\,,
\ee
which yields the vector potential
\be
  \mathbf{A(\mathbf{r})} = \frac{1}{\pi}\int_0^{2 \pi}\dd\phi \frac{-a \sin (\phi)\hat{\boldsymbol{x}}+ b \cos (\phi)\hat{\boldsymbol{y}}}{(x-a \cos (\phi))^2+ (y-b \sin (\phi))^2 +z^2}\,.
\ee
The above integrals can still be done analytically in certain cases \cite{Pinto:2022bho}, however, the expressions are long and are not particularly illuminating so we will not transcribe them here. After taking the curl, and transforming the result into AdS, the resulting magnetic field is divergenceless, as it should be by construction. However, upon inspecting the norm, one can observe that it does not saturate the bound $V=1$ anywhere but at the boundary of AdS, hence it does not define a bulk surface homologous to the entangling region. This means that the modified Biot-Savart law does not generally hold for arbitrary shapes and does not give a valid bit thread configuration.  This does not contradict the maximization of the flux since our previous analysis only showed that this prescription gives a local maximum, not a global one.

This raises a number of questions: does, perhaps, a more general Biot-Savart law apply? Our main assumptions in postulating \eqref{eq:magneticfield} were linearity, homogeneity, and isotropy with respect to the source. Given that the principle of superposition does not apply to bit thread configurations (as entanglement entropy is inherently nonlocal), any linear law must fail for at least some configurations. However, we cannot completely rule out a more general law, perhaps if drop some of our fundamental assumptions discussed above. It is also interesting to study what happens for infinitesimal perturbations. The failure to fulfill the norm bound excludes the possibility of magnetic bit threads for finite perturbations around symmetric configurations. However, it is possible that the magnetic formula reproduces the right result for infinitesimal perturbations up to some order, as we will explicitly verify below. Going the other way around, we would like to understand if geodesics can always be a good candidate for the integral curves of a bulk flow, regardless of the shape of the entangling region, and if so, what would it imply for the modifications to the prescription found using magnetic fields. In the next section, we will explore these issues in a systematic way, namely, by developing a consistent perturbative expansion for bit threads involving generic entangling regions.

%%%%%%%%%%%%%%%%%%%%%%%%%%%%%%%%%%%%%%
%%%%%%%%%%%%%%%%%%%%%%%%%%%%%%%%%%%%%%
\section{Exploring shape deformations}
\label{sec:deform}
%%%%%%%%%%%%%%%%%%%%%%%%%%%%%%%%%%%%%%
%%%%%%%%%%%%%%%%%%%%%%%%%%%%%%%%%%%%%%

As shown at the end of the previous section, the prescription based on a local magnetic field does not work for general entangling regions. The goal of this section is to explore a different approach and figure out what the corrections to the magnetic field picture may look like, at least perturbatively close to the cases that are correctly reproduced with our prescription. Our eventual objective is to develop a systematic way to compute bit thread configurations for generic entangling regions. 

\subsection{Perturbations of RT surfaces}

As preparation for considering perturbed thread configurations, we will first analyze how to compute general perturbations of RT surfaces in asymptotically AdS$_{d+1}$ spaces.\footnote{While we ultimately want to consider the AdS$_4$ case, most of the results of this section apply to any dimensionality $d\geq3$. We will fix the dimension of spacetime only at the end of the section when looking at concrete examples and making a connection with `magnetic' bit threads. Also, notice that shape deformations in AdS$_3$ are trivial, as the only possible deformations of a 1-dimensional interval amount to a change of length.} To make everything as surface independent as possible, we will use Gaussian normal coordinates adapted to the surface of interest. We start by fixing a coordinate system on the RT surface $y^i$, $i=\{1,\ldots,d-1\}$, and then we extend them to a neighborhood by parallel transporting them along geodesics orthogonal to the surface. This defines a set of coordinates $\{w, y^i\}$ where $w$ is the affine parameter along the normal geodesics. In these coordinates, the spatial part of the bulk metric takes the form
\begin{equation}
        \dd s^2 = \dd w^2 + \gamma_{ij} \dd y^i \dd y^j\,,
\end{equation}
while the embedding of the RT surface is simply given by $w=0$. Gaussian normal coordinates will always exist in a neighborhood of the surface, however, depending on the surface they may or may not be globally well-defined. 

\begin{figure}[t!]
\centering
 \includegraphics[width=1.7in
 ]{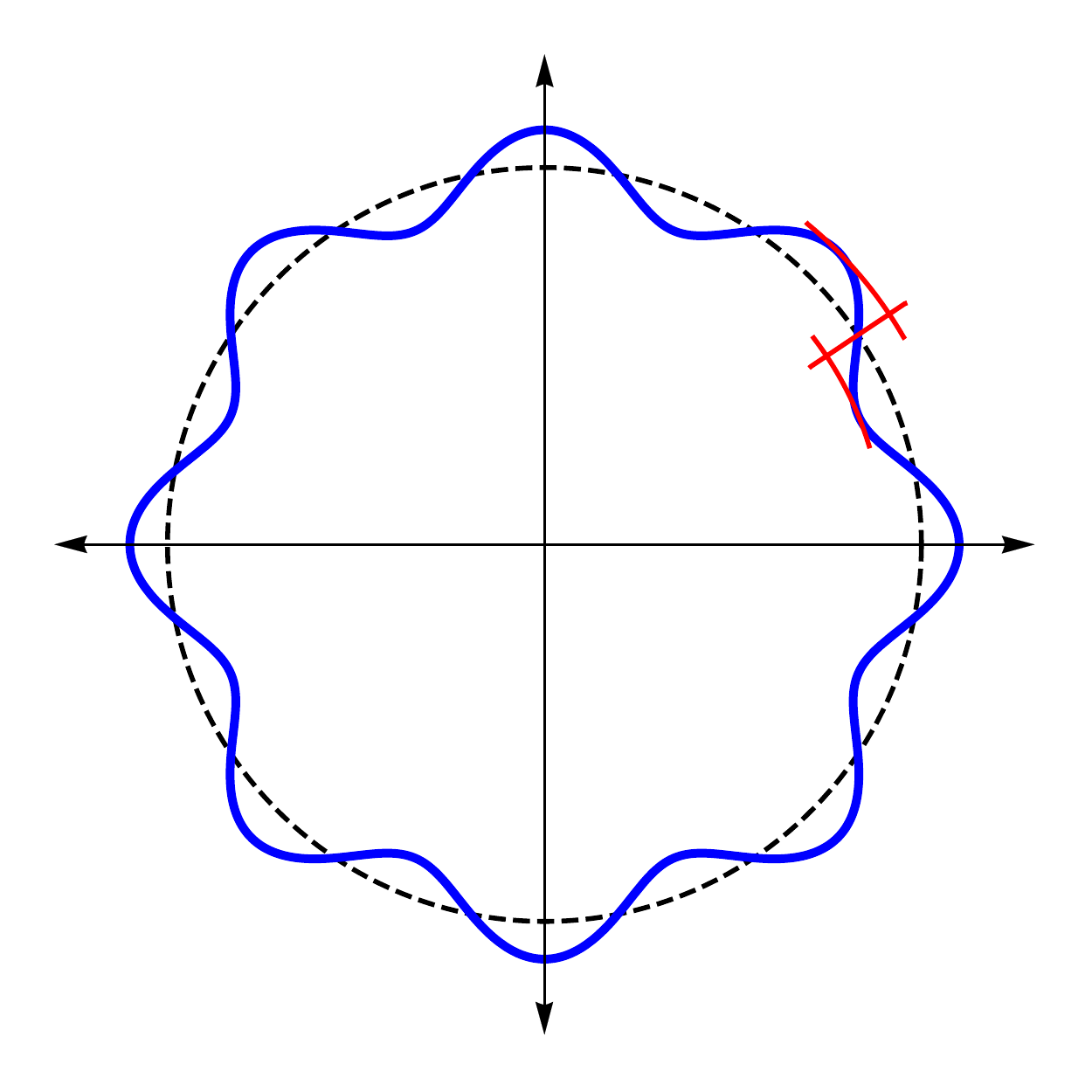}\hspace{1cm}
  \includegraphics[width=3.in
 ]{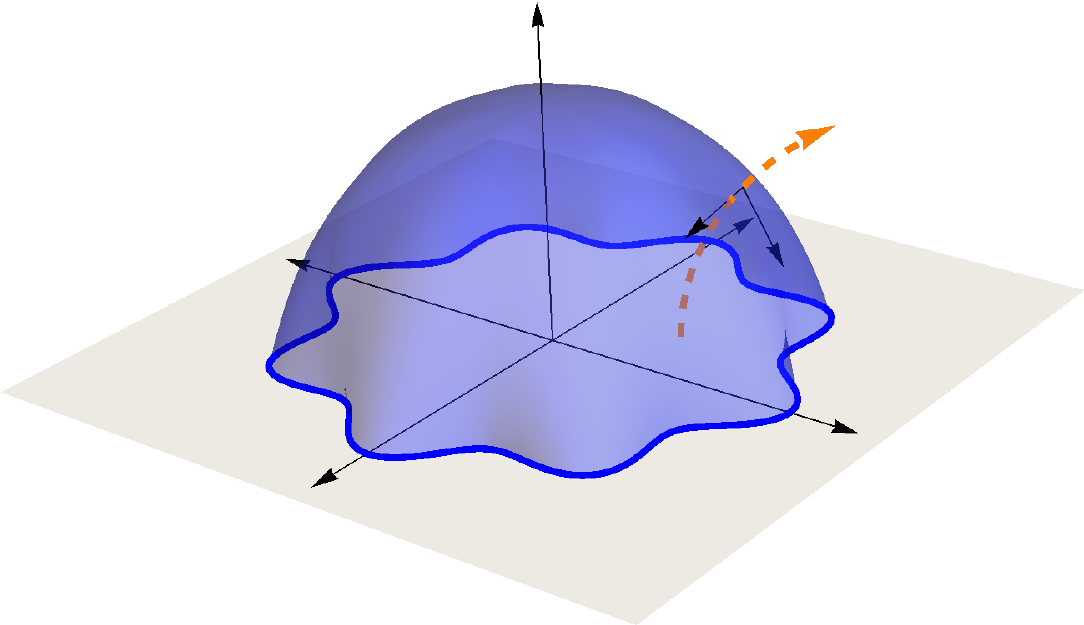}
 \begin{picture}(0,0)
\put(-319,120){{\small $x_2$}}
\put(-258,59.5){{\small $x_1$}}
\put(-274,80){{\scriptsize{\color{red}$\sim\!\lambda$}}}
\put(-280,21){{\small{\color{blue}$\partial \mathcal{A}$}}}
\put(-42,33){{\small $x_1$}}
\put(-65,87){{\small $x_2$}}
\put(-115.5,127){{\small $z$}}
\put(-92,73){{\small $y_1$}}
\put(-63,68){{\small $y_2$}}
\put(-52,98){{\small{\color{orange}$w$}}}
\put(-105,21){{\small{\color{blue}$\partial \mathcal{A}$}}}
\end{picture}
 \caption{Generic shape deformation of a circle and the corresponding RT surface in a time slice of AdS$_4$. In Gaussian normal coordinates, $\{w,y^i\}$, the perturbed RT surface is given implicitly by the equation $w- \lambda h (y^i)=0$. \label{Fig:Shapes}}
\end{figure}
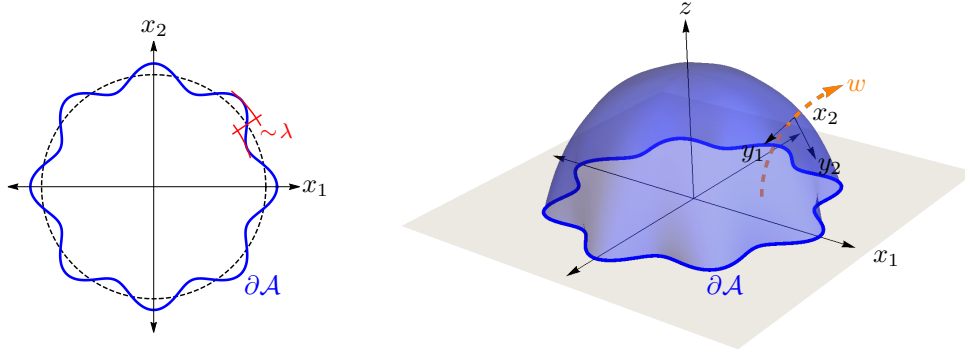

Next, we consider a slightly deformed RT surface, e.g., corresponding to a perturbed entangling region. The equation describing the new surface will be given by $w- \lambda h (y^i)=0$ for some small $\lambda$. See Figure \ref{Fig:Shapes} for an illustrative example.

The area functional for this surface is given by
\begin{align}
\text{Area}&= \text{Area}_0 + \frac{\lambda^2 }{2}\int \dd^{d-1} y \sqrt{\gamma} \left(\gamma^{ij} \partial_i h \partial_ j h + \frac{1}{2\sqrt{\gamma}} h^2 \partial_ w^2 \sqrt{\gamma}\right)+ \mathcal{O}(\lambda^3)\,.
\end{align}
Note that the unperturbed surface already minimizes the functional at zero order and, as a consequence, the first-order variation has to vanish. This implies
\begin{equation}
\label{eq:condition_dwgamma}
    \frac{1}{\sqrt{\gamma}}\partial_ w \sqrt{\gamma}|_{w=0}= \gamma^{ij}\partial_ w \gamma_{ij}|_{w=0}=0\,.
\end{equation}
The Euler-Lagrange equation associated with the first correction is
\begin{equation}
    \partial_ i \left( \sqrt{\gamma} \gamma^{ij} \partial_ j h \right)=h \partial_ w ^2 \sqrt{\gamma}\quad \Rightarrow\quad \nabla^2 h= h  \frac{\partial_w^2 \sqrt{\gamma}}{\sqrt{\gamma}}\,.
\end{equation}

\subsubsection{Geometry of AdS in Gaussian normal coordinates}
So far, we have looked at RT surfaces on arbitrary backgrounds. To make further progress on our problem, we will concentrate on pure AdS$_{d+1}$ spaces.\footnote{Or rather, Euclidean AdS$_{d}$ spaces, as we are considering only the spatial part of the geometry.} This will allow us to determine the relation between $w$ and $\gamma$, which will simplify our subsequent analysis. 

We use the fact that pure AdS is a maximally symmetric space to write the Riemann tensor in terms of the metric as follows
\begin{equation}\label{eq:RiemannAdS}
    R^i_{\;w w j}= \gamma^{ik} \gamma_{kj}=  \delta^i_ j\,.
\end{equation}
On the other hand, we can explicitly compute this component of the Riemann tensor by taking covariant derivatives of the metric. For this calculation we need 
the following expressions for the Christoffel symbols in Gaussian normal coordinates,
\begin{align}
\Gamma^w_{ij}&=- \frac{1}{2} \partial_ w \gamma_{ij} \label{eq:christoffel_1}\,,\\
\Gamma^i_{jw}&=\frac{1}{2} \gamma^{ik} \partial_w \gamma_{jk}
\label{eq:christoffel_2}\,,\\
\Gamma^i_{jk}&=\frac{1}{2} \gamma^{i l} \left(  \partial_ j \gamma_{k l}+\partial_ l \gamma_{kj}-\partial_ l g_{jk} \right)\,,
\label{eq:christoffel_3}
\end{align}
with the remaining ones vanishing. After some algebra, we arrive at
\begin{align}
    \label{eq:Riemann_normal_coordinates}
    R^i_{\; w w j}&=\frac{1}{2}\partial_w\left( \gamma^{ik}\partial_w \gamma_ {kj}\right) + \frac{1}{4} \gamma^{il}\partial_ w\gamma_ {lk} \gamma^{km} \partial_ w \gamma_ {mj}\,.
\end{align}
Combining \eqref{eq:RiemannAdS} and \eqref{eq:Riemann_normal_coordinates} we can now obtain a matrix equation for $\gamma$
\begin{equation}
\label{eq:A_diff_equation}
    \partial_w \left( \gamma^{-1} \partial_w \gamma \right) + \frac{1}{2} \gamma^{-1} (\partial_w  \gamma) \gamma^{-1} (\partial_w \gamma) = \partial_w M + \frac{1}{2} M^ 2= 2\,,
\end{equation}
where we have defined $M\equiv \gamma^{-1} \partial_w \gamma$. Equation \eqref{eq:A_diff_equation} can be integrated assuming that there exists a matrix $U$ independent of $w$ that diagonalizes $M(w)$ as $M(w)= U^{-1} D(w) U$. This gives the following solution for $M$,
\begin{equation}
\label{eq:dgamma_gamma_z}
    M(w)= 2\, U^{-1} \tanh \left( \tanh^{-1} \left( \frac{ D(0)}{2} \right) + w \right) U\,.
\end{equation}
Since this is a first-order differential equation and we have found a solution parametrized by an arbitrary initial matrix, this must be the general solution. The structure of the differential equation already tells us that $M$ commutes with its derivative, thus it is no surprise that a common diagonalization for all $w$ exists. Finally, for $\gamma$ we obtain 
\begin{align}\label{gammaG}
    \gamma(w)&=  \gamma(0) \left( \cosh ( w  ) +\frac{ \gamma^{-1}(0) \partial_w \gamma(0)}{2}\sinh ( w  )  \right)^2 \,.
    \end{align}
This solution implies that once we fix the induced metric on the RT surface and its normal derivative we have all the information we need to find the metric everywhere else. As a consequence, we have been able to reduce drastically the amount of geometric information we need to provide. Using this solution, we are able to compute $\frac{\partial_ w ^2 \sqrt{\gamma}}{\sqrt{\gamma}}$ in terms of $\gamma_{ij}(0)$ and $\partial_w \gamma_ {ij}(0)$. The equation for $h$ then becomes 
\begin{equation}
\label{eq:flow_h}
    \nabla^2 h= (d-1)h - \frac{h }{4} \Tr \left[  \gamma^{-1} \partial_w \gamma \gamma^{-1} \partial_w \gamma \right]\,,
\end{equation}
where $d$ is the dimensionality of the boundary spacetime (e.g., $d=3$ for AdS$_4$). This equation applies to any RT surface irrespective of its symmetries. The only information about the original RT surface is contained in the metric and its normal derivative.

\subsection{Flow equations for RT surfaces}

Equation \eqref{eq:flow_h} gives us $h$ for any $\gamma$, thus, if we know how $\gamma$ evolves under a set of consecutive infinitesimal perturbations, we could in principle find the first-order perturbations along the whole evolution and then integrate them to obtain an expression for the final RT surface. In this section, we will obtain a flow equation describing such an evolution.

Our first step is to adapt our coordinates so that they stay normal to the perturbed RT surface at any point in the evolution. We will denote all quantities referring to the old RT surface by a subscript $0$ and the ones referring to the perturbed RT surface by a subscript $\lambda$. By definition $\nabla w_ \lambda =\hat{n}$ where the normal vector $\hat{n}$ is given by
\begin{equation}
\hat{n}= \left(1, -\lambda \gamma^{ij}_0 (w_0=0) \frac{\partial h }{\partial y_0^j}   \right)\,.
\end{equation}
We will now define $y^{i}_\lambda$ by parallel transporting it back along the $\lambda$ evolution. In particular, when we parallel transport a point with coordinates $w_{\lambda}=0$ and ${y_{\lambda}}^i=a^i$ to $w_0=0$ along the normal geodesic to the new RT surface we obtain $y_0^i=a^i$. The corresponding change of coordinates can be expanded in $\lambda$ as
\begin{align}
    w_0(w_{\lambda},y^i_{\lambda})&=w_{\lambda} + \lambda \tilde{w}(w_{\lambda},y^i_{\lambda})+ \mathcal{O}(\lambda^2)\,,\\
    y_0^i(w_\lambda, y^i_{\lambda}) &=y^i_{\lambda} + \lambda \tilde{y}^i(w_{\lambda},y^i_{\lambda})+ \mathcal{O}(\lambda^2)\,.
\end{align}
By definition $w_\lambda$ has to be an affine parameter for a geodesic normal to the new RT surface, thus we need to solve the geodesic equation. Using the Christoffel symbols \eqref{eq:christoffel_1}-\eqref{eq:christoffel_3} we find
\begin{equation}
\frac{\dd ^2 w_0}{\dd {w_{\lambda}}^2 }- \partial_ {w_0}  {\gamma_0}_{ij} \frac{\dd y_0^i}{\dd w_{\lambda}} \frac{\dd y_ 0^j}{\dd w_{\lambda}}=\lambda \frac{\dd ^2 \tilde{w}}{\dd {w_{\lambda}}^2 }+ \mathcal{O}( \lambda^2)=0\;\quad\Rightarrow\;\quad \tilde{w}=C_1+ w_{\lambda} C_2+ \mathcal{O}( \lambda^2)\,.
\end{equation}
The boundary conditions tell us that $w_0(w_{\lambda}=0)= \lambda h$, and $\left.\frac{\dd w_0}{\dd w_{\lambda}}\right|_{w_{\lambda}=0}=1 $. Therefore
\begin{equation}
\label{eq:flow_w}
    w_0= w_{\lambda} + \lambda h(y_{\lambda}^i) + \mathcal{O}(\lambda^2)\,.
\end{equation}
Repeating for the transverse coordinates, we obtain
\begin{align}
\frac{\dd ^2 y_ 0^i}{\dd {w_{\lambda}}^2 }+ \Gamma^i_{jk} \frac{\dd y_ 0^j}{\dd w_{\lambda}} \frac{\dd y_ 0^k}{\dd w_{\lambda}}+{\gamma_0}^{ik} \partial_{w_0} {\gamma_0}_{jk} \frac{\dd y_0^j}{\dd w_{\lambda}} \frac{\dd w_0}{\dd w_{\lambda}}=&\nonumber\\
\lambda\frac{\dd ^2 \tilde{y}^i}{\dd {w_{\lambda}}^2 }+\lambda{\gamma_0}^{ik} \partial_{w_0} {\gamma_0}_{jk} \frac{\dd \tilde{y}^j}{\dd w_{\lambda}}+ \mathcal{O}(\lambda^2)=& \,\,0\,,
\end{align}
with boundary conditions $y^i_0(w_0=0)= y_{\lambda}^i$ and $\left.\frac{\dd y_0^i}{\dd w_{\lambda}}\right|_{w_{\lambda}=0}=-  \lambda {\gamma_0}^{ij} \partial_j h $. The solution is
\begin{equation}
\label{eq:flow_yi}
    y_0^i =y_{\lambda}^i - \lambda\int_ 0 ^{w_{\lambda}} \dd s {\gamma_0}^{ij}(s,y_{\lambda}^k) \partial_ j h(y_{\lambda}^k)+ \mathcal{O}(\lambda^2)\,,
\end{equation}
as can be checked by direct substitution. The transverse metric on the new RT surface yields
\begin{align}
\label{eq:flow_gamma}
{\gamma}^\lambda_ {ij} (w_{\lambda}, y_{\lambda}^i)&=  \gamma_{ij}+  \lambda h \partial_w \gamma_{ij} - \lambda \partial_{l}\gamma_ {ij}\int \dd w \gamma^{lk} \partial_k h  \nonumber\\
&\phantom{=}-  \lambda \gamma_{ik}\int \dd w \partial_{j} \left( \gamma^{kl} \partial_ l h \right)-\lambda \gamma_{jk}\int \dd w \partial_{i} \left( \gamma^{kl} \partial_ l h \right)+ \mathcal{O}(\lambda^2)\,.
\end{align}
For the sake of lightening the notation, we have suppressed the indices in the final expression since they all refer to the old RT surface so there is no room for confusion.

Equation \eqref{eq:flow_gamma} can now be recast as a differential equation for $\gamma$,
 \begin{equation}
\frac{\dd  \gamma_ {ij}}{\dd \lambda }  = h \partial_w \gamma_{ij} -  \partial_{l}\gamma_ {ij}\int \dd w \gamma^{lk} \partial_k h-  \gamma_{il}\int \dd w \partial_{j} \left( \gamma^{lk} \partial_ k h \right) -\gamma_{jl}\int \dd w \partial_{i} \left( \gamma^{lk} \partial_ k h \right)\,.
 \end{equation}
Further, since we only need $\gamma_{ij}(0)$ and $\partial_w \gamma_{ij}(0)$ we can simply evaluate at $w=0$ to obtain
\begin{equation}
\label{eq:evol_2}
 \frac{\partial \gamma_{ij}  }{\partial \lambda }   = h \partial_w \gamma_{ij}\,.
\end{equation}
As for the normal derivative, we get
\begin{align}
\label{eq:flow_dgamma}
    \frac{\partial  }{\partial  \lambda}  \partial_w \gamma_{ij}&=2 h \gamma_ {ij} + \frac{h }{2} \partial_ w \gamma_ {ik} \gamma^{kl} \partial_ w \gamma_{lj} - 2\nabla_ i \nabla_ j h\,.
\end{align}
These two equations \eqref{eq:evol_2} and \eqref{eq:flow_dgamma}, together with \eqref{eq:flow_h}, define a system of non-linear equations that determine the flow of $h$ and $\gamma_{ij}$ under \emph{arbitrary} evolution in $\lambda$. Note that an exact solution of these equations gives the exact position of the RT surface for a finite value of $\lambda$ and there is no approximation involved even if we have expanded everything to first order. One can think of $h$ as the ``velocity'' of the RT surface. The equation \eqref{eq:flow_h} gives us an exact equation for it supposing we know $\gamma$ and $\partial_w \gamma$. Then, due to the maximal symmetry of AdS we can reconstruct the whole metric from information just at a single initial surface, and thus we can set up equations \eqref{eq:evol_2} and \eqref{eq:flow_dgamma} which will keep track of how the metric changes when deforming the RT surface continuously. For this procedure to work, symmetry is absolutely essential since without it there is no possibility of arriving at a well-posed initial value problem. In other words, without symmetry, the information of a single surface would only determine the properties of space around that surface.

It is also important to point out that the change in coordinate systems is already built into these equations. Once a solution is found, if we want to recover an expression in terms of the original coordinate system, we would need to solve \eqref{eq:flow_w} and \eqref{eq:flow_yi} in differential form, namely
\begin{align}
\label{eq:flow_coordinates1}
    \frac{\dd w}{\dd \lambda}&= - h(y^i)\,, \\
\label{eq:flow_coordinates2}    \frac{\dd y^i}{\dd \lambda}&=\int_0^{w} \gamma^{ij} \partial_j h \;\dd w\,,
 \end{align}
 with $\gamma$ and $h$ already determined by a known solution to equations \eqref{eq:flow_h}, \eqref{eq:evol_2} and \eqref{eq:flow_dgamma}. The original coordinate system would show up as integration constants of \eqref{eq:flow_coordinates1}-\eqref{eq:flow_coordinates2} and the resulting relation would then need to be inverted.

\subsubsection{Higher order perturbations from the flow equations}

The flow equations, \eqref{eq:flow_h}, \eqref{eq:evol_2} and \eqref{eq:flow_dgamma}, describing a set of consecutive infinitesimal perturbations, can be in principle used to find arbitrarily higher order perturbations to the original quantities. In this section, we will outline such a procedure. For concreteness, we will focus on perturbations to a spherical entangling region, but the same methodology would apply if we choose to perturb around another region of interest.

The first step is to find $\gamma$ in Gaussian normal coordinates. In fact, we only need the expansion in $w$ up to order $\mathcal{O}(w)$ but we will obtain the full transformation as it will be useful at a later stage. Without loss of generality, we will consider the case of a  sphere of radius $R$ centered at the origin. The normal geodesics in the affine parameterization are then
\begin{align}
z&=\frac{\sqrt{R^2-r_s^2} R}{r_s} \sech \left(w-\text{arcsech} \frac{r_s}{R}\right)\,,\\\label{eq:geodesic_r}
r&=r_s+ \frac{R^2-r_s^2}{r_s} + \frac{\sqrt{R^2-r_s^2} R}{r_s} \tanh \left(w-\text{arcsech} \frac{r_s}{R}\right)\,,
\end{align}
where $r_s\in [0,R]$ is a parameter dictating the point where the geodesic intersects the RT surface. We can invert $r_s$ in terms of $r$ and $z$ as
\begin{equation}
\label{eq:change_coordinatesr}
    r_s= \frac{2 R^2 r}{R^2+ r^2 +z^2}\,.
\end{equation}
We can also compute $\gamma$ explicitly in terms of the variable $r_s$. This gives the following line element for the transverse space
\begin{equation}
  ds_\perp^2\equiv\gamma_{ij}dx^idx^j= \frac{R^2}{(R^2- r_s^2)^2}\dd r_s^2 + \frac{r_s^2}{(R^2-r_s^2)}\dd \Omega^2_{d-2}\,.
\end{equation}
Thus, the equation for $h$ becomes
\begin{equation}
     \frac{1}{\sqrt{\gamma}} \partial_{r_s} \left( \sqrt{ \gamma} \frac{(R^2-r_s^2)^2}{R^2} \partial_{r_s} h  \right)  + \frac{R^2 -r_s^2}{r_s^2 }\nabla_{S_{d-2}} h =(d-1)h\,.
\end{equation}
It is convenient to expand the angular part in spherical harmonics, which fulfills
\begin{equation}
    \nabla_{S_{d-2}} Y_{jm}=-j(j+d-3) Y_{jm}\,,
 \end{equation}
with $m=\left\{ m_1,\dots m_{d-3} \right\}$. For each $j$ we can now find an equation for the radial part. Defining the differential operator
\begin{equation}
    L f(r_s) \equiv \frac{1}{\sqrt{\gamma}} \partial_{r_s} \left( \sqrt{ \gamma} \frac{(R^2-r_s^2)^2}{R^2} \partial_{r_s} f(r_s)  \right)\,,
\end{equation} 
we find
\begin{equation}
    L f_j(r_s)   -j(j+d-3) \frac{R^2 -r_s^2}{r_s^2 } f_j(r_s) =(d-1)f_j(r_s)\,,
\end{equation}
which can be solved to obtain
\begin{align}
     f_j(r_s)= \frac{\sin (\frac{d \pi}{2}) \Gamma ( 1- \frac{d }{2}) \Gamma ( \frac{d+j-1}{2}) \Gamma ( \frac{d+j}{2})   }{\pi \Gamma ( \frac{d+2j-1}{2}) \sqrt{R^2-r_s^2}}\left( \frac{r_s}{R} \right)^j \,_2F_1 \left( \tfrac{j-1}{2},\tfrac{j}{2},\tfrac{d-1+2j}{2}, \tfrac{r_s^2}{R^2} \right)\,.
\end{align}
To obtain this solution, we have required that there should be no divergences at the location of the RT surface, and we have normalized the solution such that
\begin{equation}
    \lim_{r_s\to R}\sqrt{R-r_s^2}f_j(r_s)=C_3 \,_2F_1 \left( \tfrac{j-1}{2},\tfrac{j}{2},\tfrac{d-1+2j}{2}, 1 \right)=1\,.
\end{equation}
The full expression for $h$ is
\begin{equation}
\label{eq:h_rs_phii}
    h(r_s, \varphi^{(i)})=\sum_{j,m} a_{jm} Y_{jm}(\varphi^{(i)}) f_j(r_s) \equiv \sum_{jm }a_{jm} h_{jm}\,.
\end{equation}
Note that the $a_{jm}$ coefficients have to be fixed from the boundary conditions. However, there are some technicalities to this that will be discussed at a later stage.

Equation \eqref{eq:h_rs_phii} agrees with the leading-order result obtained in \cite{Mezei:2014zla}. However, with the flow equations \eqref{eq:evol_2} and \eqref{eq:flow_dgamma}, we can now obtain a solution at arbitrarily high orders. This is done by expanding $h$ as a Taylor series in $\lambda$,
\be
h(\lambda,r_s, \varphi^{(i)})=h(\lambda=0,r_s, \varphi^{(i)})+\sum_{n=1}^{\infty}\frac{\lambda^{n}}{n!}h^{(n)}(\lambda=0,r_s, \varphi^{(i)})\,.
\ee
We have already found the first term in the expansion. To go beyond this order, we differentiate eq. \eqref{eq:flow_h} $n$ times with respect to $\lambda$,
\begin{align}\label{eqhn}
\nabla^2 h^{(n)}&= (d-1) h^{(n)}-\frac{h^{(n)}}{4} \Tr \left[ \gamma^{-1} \partial_w \gamma \gamma^{-1} \partial_w \gamma \right] +F_n(h^{(n-1)},\dots, h, \gamma,\partial_w \gamma)\,,
\end{align}
where $F_n$ is some function that can be easily determined. This function receives contributions from derivatives of $\gamma$ in the trace term and in the Laplacian with respect to $\lambda$ which we remove using equations (\ref{eq:evol_2}) and (\ref{eq:flow_dgamma}).

We will now proceed by induction. The first term, $h(\lambda=0)$, is known from eq. \eqref{eq:h_rs_phii}. If we assume we know all $h^{(k)}(\lambda=0)$ for $k<n$ as explicit functions of $r_s$ and $\varphi^{i}$, then, we can find $h^{(n)}(\lambda=0)\equiv g(r_s,\varphi^i)$ by solving the equation 
\begin{equation}
\label{eq:non-homogeneous_part}
\nabla^2 g -(d-1) g = F_n(r_s, \varphi^{i})\,,
\end{equation}
where we used the fact that the trace term in (\ref{eqhn}) vanishes at $\lambda=0$ for a maximally symmetric surface. 
We can separate the solution into homogeneous and inhomogeneous parts. The solution to the homogeneous equation will have the form of \eqref{eq:h_rs_phii}, with coefficients $a_{jm}^{(n)}$ fixed by the higher order corrections to $\partial\mathcal{A}$. The inhomogeneous part of the solution can be obtained using Green functions. The Green function of interest must be a solution to the equation
\begin{equation}
\label{eq:GreenFunction}
\nabla^2 G(y,y') - (d-1) G(y,y') = \frac{1}{\sqrt{\gamma}} \delta^{(d-1)}(y-y')\,,
\end{equation}
with vanishing boundary conditions on $\partial\mathcal{A}$. Since the unperturbed RT surface is maximally symmetric, we can assume that the propagator only depends on the geodesic distance between $y$ and $y'$, $\sigma(y,y')$, so that $G(\sigma)$ must satisfy
\begin{equation}
      \frac{\dd^2 G }{\dd \sigma^2} \nabla_ i \sigma\nabla^i \sigma + \frac{\dd G}{\dd \sigma} \nabla^2 \sigma  - (d-1) G = \frac{1}{\sqrt{\gamma}} \delta^{(d-1)}(y-y')\,.
  \end{equation}
More explicitly, for a spherical entangling region, the unperturbed RT surface is simply an AdS$_{d-1}$ space,  just in a non-standard coordinate system\footnote{The RT surface is Euclidean AdS$_2$ when the bulk is AdS$_4$. Our coordinate system covers a single boundary of this AdS$_2$.}. Thus, we can actually use the Dirichlet propagator for an AdS space in Euclidean signature. Upon doing so, we obtain
\begin{equation}\label{eq:generalhn}
h^{(n)}(\lambda=0,r_s,\varphi^i) = \sum_{jm} a^{(n)}_{jm} f_j(r_s) Y_{jm}(\Omega) + \int \sqrt{\gamma}\dd r_s' \dd \Omega'  G(r_s, \Omega,r_s' , \Omega') F_n (r_s', \Omega')\,,
\end{equation}
giving us an explicit expression for $h^{(n)}(\lambda=0)$. We are able to do this for arbitrarily high $n$, by induction. Thus, we can find higher-order shape perturbations just by taking derivatives and performing integrals, without the need of solving any equation beyond the first order. 

\subsubsection*{General structure and interpretation}

In practical examples, it is simpler to expand eq. \eqref{eq:non-homogeneous_part} in Fourier modes and solve the one-dimensional boundary value problem for each mode. However, the general solution \eqref{eq:generalhn} reveals manifestly the structure of the problem. In particular, we note that \eqref{eq:generalhn} can be naturally interpreted in terms of a diagrammatic expansion, as follows. 

First, we write the homogeneous part as an integral with a fixed kernel in the form
\begin{equation}
    h^{(n)}_{hom}(r_s,\varphi^i)= \int \dd r'_s \dd \Omega' \delta(r_s-r_s') \delta(\varphi_i - \varphi_i') h^{(n)}_{hom}(r_s',\varphi_i')=\int_{r_s'=R} \sqrt{\gamma} \dd \Omega'  h^{(n)}_{bc}(\varphi_i')\partial_{r'_s} G\,,
\end{equation}
where $h_{bc}^{(n)}(\varphi_i)$ are functions that parametrize $\partial\mathcal{A}$ at different orders in $\lambda$ and, thus, determine the boundary conditions for $h^{(n)}$. To obtain the second equality we have used equation \eqref{eq:GreenFunction} and integrated by parts taking into account the boundary conditions for $G$. The expression for the kernel in this expression is actually just the definition of the familiar bulk-to-boundary propagator, up to the adequate power of $R-r_s$ to take care of the volume divergences associated with the AdS metric. Notice also that the $F_n$ functions depend only on $h$ and its derivatives. Thus, proceeding by induction, as above, we should be able to replace them with other expressions in terms of propagators, possibly acting with some factors of the momentum. Overall, they will define the vertices of some interacting theory such that we can write the net result for $h$ as a sum over Witten-like diagrams. 

\begin{figure}[t!]
\centering
 \includegraphics[width=3.5in
 ]{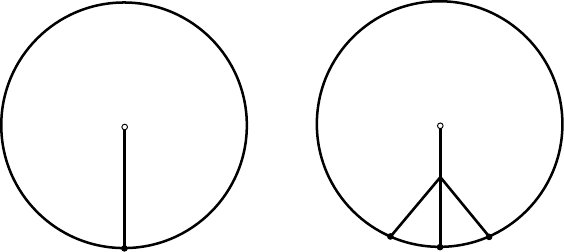}
 \caption{Witten diagrams contributing to the modification of the RT surface up to second order in $\lambda$. The white dot represents a delta function that fixes the value of $h$ at the specific point we want to compute. The boundary conditions for the desired RT surface are inserted at the boundary sources. The left (right) figure represents the homogenous (inhomogenous) contributions where the quartic vertex in the bulk arises from (\ref{eq:f2}).}\label{Wittendiagrams}
\end{figure}

As an example, consider the result for $h$ up to order $\mathcal{O}(\lambda^2)$. It is easy to check that $F_1$ vanishes, while $F_2$ takes the form
\begin{align}
\label{eq:f2}
    F_2=2 h (h \delta^j_i - \nabla_i \nabla^j h)(h \delta^i_ j-\nabla_j\nabla^i h)+ 2 \nabla_j \left( h( h \delta^j_ k -\nabla_k\nabla^j h)\nabla^k h \right)\,.
\end{align}
Then, up to $\mathcal{O}(\lambda^2)$, there are contributions from the diagrams shown in Figure \ref{Wittendiagrams}. Notice that $h^{(n)}$ appears multiplied by $\frac{\lambda^n}{n!}$ in the expression for $h$, however, these factors can be absorbed in the definition of the vertices. Further, the boundary conditions enter into the diagrams as boundary sources. Evidently, vertices with arbitrary higher powers of $h$ will appear at higher orders in $\lambda$, ultimately rendering the theory nonlocal. Neglecting these higher-order vertices we can formally resum all the contributions from the vertices with up to fourth powers with a Dyson equation of the form
\begin{gather}
\begin{aligned}
    \includegraphics[width=3in]{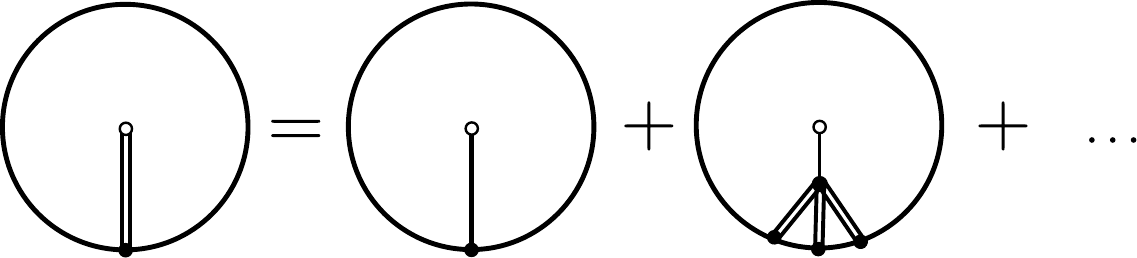}
\end{aligned}
    \raisetag{20pt}
\end{gather}
However, solving this equation does not seem to be a simplification of the problem as it is a nonlinear integral equation.

In retrospect, the fact that we obtain inhomogeneous equations when considering general perturbations implies that linearity and locality are generically broken, as the inhomogeneous parts are precisely the ones that encode the higher-point bulk interactions. Had we obtained homogeneous equations only, this would have implied the full solution is captured by the first of the Witten diagrams alone. Such a term could be entirely encapsulated by a classical (magnetic-like) field. However, the inhomogeneous nature of the equations tells us that such a prescription ultimately breaks down. In contrast, the structure of the equations suggests the actual picture resembles more a version of nonlocal quantum electrodynamics. We will come back to this point when discussing shape perturbations to bit threads.

\subsubsection{Boundary conditions}

So far we have not addressed how to deal with the boundary conditions. Naively one would expect to impose them by requiring
\begin{equation}
    \lim_{r_s\to R} r(w= w|_{RT}, r_s , \varphi^{(i)} )= r(\varphi^{(i)})\,,
\end{equation}
where $r(\varphi^{(i)})$ parametrizes the boundary of the entangling region, $\partial\mathcal{A}$. This, however, runs into a major problem. The issue is that 
the limit $r_s\to R$ will always yield $r\to R$, as can be seen from the expression of $r(w,r_s, \varphi^{(i)})$.
Moreover, any point that touches the boundary and has $r_s<R$ must necessarily have a divergence in $w$, suggesting that modifications of the boundary conditions must be non-perturbative.

It can be checked that these issues are just artifacts of the choice of coordinate system. Luckily, we can avoid these complications by introducing a UV cutoff at $z=\epsilon\ll R$ and imposing the boundary conditions on this surface. However, some caution is still necessary when considering the expansion in $\lambda$. 

The first step is to figure out the value of $w_0$ at the RT surface denoted by $w_0|_{RT}$ at a given order in $\lambda$. Note this is not $h$, in general. It is fixed by the solution to equations \eqref{eq:flow_w} and \eqref{eq:flow_yi}. In particular, to $\mathcal{O}(\lambda^3)$, we find
\begin{align}
     w_ \lambda &= w_0 - \lambda h - \frac{\lambda^2}{2} \left( h^{(1)}+ \partial_i h \partial_ j h \gamma^{ij} \tanh w_0 \right) \nonumber\\
     &\phantom{=} - \frac{\lambda^3}{6} \left( h^{(2)} + 2 \partial_ i h^{(1)} \partial_j h \gamma^{ij} \tanh w_0 - \partial_ i h \partial_j h \gamma^{ij} \frac{h}{\cosh^2 w_0}\right) + \mathcal{O}(\lambda^4)\,.
 \end{align}
The deformed RT surface is located at $w_ \lambda=0$, which can be inverted order by order to obtain
\begin{equation}
    w_0|_{RT}= \lambda h + \frac{\lambda^2}{2} h^{(1)}+ \frac{\lambda^3}{6} \left(h^{(2)}- 4 h \partial_i h \partial_j h \gamma^{ij}\right)+\mathcal{O}(\lambda^4)\,.
\end{equation}
This expression tells us the value of  $w$ for some given $r_s$ and $\varphi^{(i)}$. The boundary condition at $z=\epsilon$ in turn fixes $r_s$ for any given $\varphi^{(i)}$. This last condition yields 
\begin{equation}
\label{eq:z=epsilon}
   \epsilon= \frac{\sqrt{R^2- r_s^2}R}{r_s}\sech \left(w_0|_{RT} -\text{arcsech} \frac{r_s}{R} \right)\,,
\end{equation}
which can be solved for under the assumption that $r_s$ is close to $R$.
Then, we expand $r(w=w|_{RT}, r_s= r_s|_{z=\epsilon}, \varphi^{(i)})$ in $\lambda$ and match the Fourier coefficients to fix all the $a_{jm}^{(n)}$ order by order. These coefficients will end up depending on the cutoff and, in fact, divergences will appear at higher orders in $\lambda$. The result of the expansion up to the second order is
\begin{align}
    r(z=\epsilon)|_{RT}&= R + \lambda\sum_{jm} a_{jm} Y_{jm} +  \frac{\lambda^2}{2} \left[ \sum_{jm} a^{(1)}_{jm} Y_{jm} + \frac{1}{R}\left( \sum_{jm} a_{jm}Y_{jm} \right)^2  \right] +\mathcal{O}(\lambda^3)\,.
\end{align}
Beyond the second order, such an expansion is best computed on a case-by-case basis since its form will depend on which types of divergences appear in $h^{(2)}$ and higher orders. 
We should point out, however, that higher order coefficients will typically pick up powers of $\frac{1}{\epsilon}$ and will therefore be divergent. Furthermore, when going to higher and higher orders in $\lambda$, these divergences seem to worsen and grow without limit. Yet, when computing the entanglement entropy, such higher-order divergences should cancel out and only the $\mathcal{O}\left( \frac{1}{\epsilon} \right)$ physical divergence should survive. We have checked this is the case up to $\mathcal{O}(\lambda^3)$.

\subsubsection{Concrete example: the ellipse\label{ss:ellipse}}

As a particular example, let us consider the case of an ellipse. The equation of the ellipse in polar coordinates is given by
\begin{align}
\label{eq:boundary_conditions_ellipse}
    r(\theta)&= \frac{b }{\sqrt{1- \varepsilon^2 \cos^2 \theta}}= b \sum_{n=0}^{\infty} \frac{(-1)^n\Gamma(\frac{1 }{2})\varepsilon^{2 n}}{2^{2n}\Gamma(n+1) \Gamma(\frac{1}{2}-n)}  \sum_{k=-n}^{n} \begin{pmatrix}
        2n\\ n- k
    \end{pmatrix} e^{2ik \theta}\,,
\end{align}
where $\varepsilon\equiv\sqrt{1-b^2/a^2}$ is the eccentricity, and $a$, $b$ are the major and minor semi axes. Note that the above equation automatically gives us the required expansion in powers of $\lambda\equiv\varepsilon^2$ in terms of spherical harmonics. The first perturbation is given by eq. \eqref{eq:h_rs_phii}. Upon imposing the boundary conditions, following the discussion of the previous subsection, we find
\begin{equation}
\label{eq:h_ellipse}
    h(\lambda=0,r_s,\theta)= \frac{1}{4\sqrt{b^2- r_s^2}} \left[1+\left(3 -  \frac{2b^2}{r_s^2}+  \frac{2b^2}{r_s ^2} \sqrt{1- \frac{r_s^2}{b^2} }^3 \right)  \cos{ 2 \theta}\right]\,.
\end{equation}
To compute the higher order corrections we need the functions $F_n(r_s,\theta)$. We will do this explicitly for $h^{(1)}(r_s,\theta)$ and $h^{(2)}(r_s,\theta)$, which require $F_1(r_s,\theta)$ and $F_2(r_s,\theta)$, respectively. As explained above, $F_1(r_s,\theta)$ vanishes generically. Using the known expression for $h$, eq. \eqref{eq:h_ellipse}, in eq. \eqref{eq:f2} yields $F_2(r_s,\theta)$ for the ellipse. The resulting expression can be used to solve for the boundary conditions as indicated in the previous subsection, and the area of the resulting RT surface can be explicitly computed. The final answer for the entanglement entropy up to $6$th order in the eccentricity becomes
\begin{equation}
\label{eq:ent_entr_ellipse}
    S= \frac{1}{4G_N} \left[\left(\frac{ 2\pi b}{\epsilon} - 2\pi\right) +  \frac{ \pi b }{2 \epsilon}\lambda +  \left( \frac{13 \pi b}{32 \epsilon}-\frac{\pi}{16} \right) \lambda^2+  \left( \frac{45 \pi b}{128\epsilon} - \frac{ \pi}{16} \right)\lambda^3 + \mathcal{O}(\lambda^4)  \right]\,.
\end{equation}
 It is easy to check that the $\frac{1}{\epsilon}$ divergence corresponds with the perimeter $P$ as expected. For more details on the calculation see Appendix \ref{ellipse_computation}. In Figure \ref{EllipseNumerics}, we have contrasted our analytical formula with the numerical results obtained for the ellipse in \cite{Fonda:2014cca}. The plotted quantity corresponds to the finite part of the entropy $\tilde{F}\equiv -(S-P/\epsilon)$, normalized as in Fig.~2 of \cite{Fonda:2014cca}. In general, we expect a very good approximation as long as the eccentricity is not too large, $\varepsilon\lesssim0.75$, consistent with our perturbative approximation.

 \begin{figure}[t!]
\centering
 \includegraphics[width=3.5in
 ]{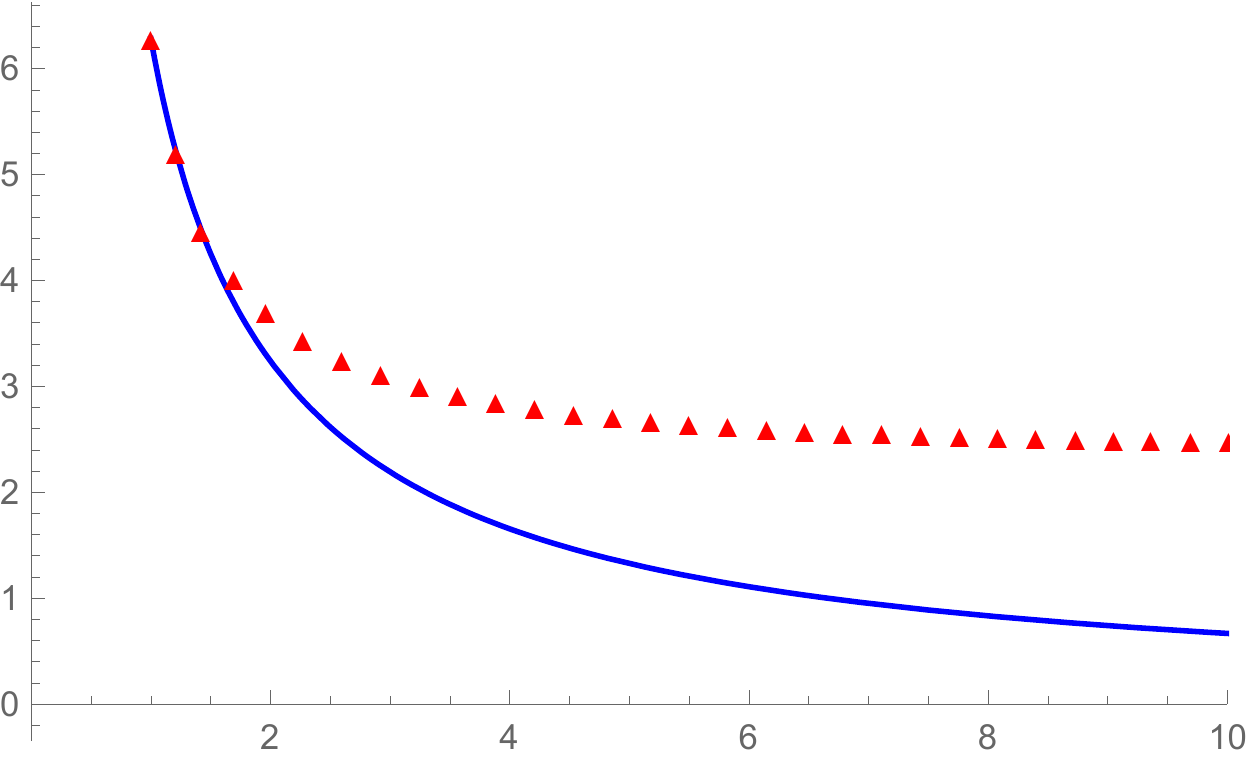}
 \put(-265,150){{\small $\tfrac{b}{a}\tilde {F}$}}
 \put(0,10){{\small $\tfrac{a}{b}$}}
 \caption{Analytic result (blue) for the finite part of the entropy $\tilde{F}\equiv -(S-P/\epsilon)$ vs. the major-to-minor axes ratio $a/b$, for an elliptical entangling region with perimeter $P$, up to $\mathcal{O}(\varepsilon^6)$.  For comparison sake, we also show the exact results (red) obtained numerically in \cite{Fonda:2014cca}. Our formula yields a good approximation up to $a/b\sim 1.5$ or, equivalently, up to $\varepsilon\sim0.75$.}\label{EllipseNumerics}
\end{figure}

\subsection{Deformed bit thread configurations from geodesics\label{sec:defgeo}}

We have developed a systematic way of computing shape perturbations for RT surfaces and now it is time to come back to bit threads.

It is well known that bit threads are highly non-unique. Intuitively, such non-uniqueness can be associated with choices of
microstates that give rise to the same amount of entanglement between the region of interest and its complement. So, different bit thread solutions codify different arrangements of EPR-type entanglement with the same macroscopic entropy. A handful of methods for computing particular thread configurations were introduced in \cite{Agon:2018lwq,Agon:2020mvu}. Although it would be interesting to understand modifications to all type of bit thread constructions under shape deformations, in this section we will focus on the case of geodesic bit threads. As discussed in section \ref{sec:magnetic}, these configurations are precisely the ones with a magnetic-like interpretation, at least for the case of the semi-infinite plane and the disk, and it would be interesting to see how the shape deformations affect this picture. 

Let us start from a known RT surface and describe the bulk spacetime using Gaussian normal coordinates. This description should be possible at least in a neighborhood of the RT surface. Following \cite{Agon:2018lwq}, we now construct a vector field representing a bit thread configuration, under the assumption that its integral lines are given by geodesics. It is easy to show that
\begin{equation}
\label{eq:bit_threads_gaussian_normal}
    V= \frac{\sqrt{\det \gamma_\lambda(w_\lambda=0)}}{\sqrt{\det \gamma_ \lambda}} \partial_{w_\lambda}\,,
\end{equation}
satisfies the desired properties. It has a unit norm at the location of the RT surface, it follows geodesics normal to it and it is divergenceless, by construction. If we can show that the norm bound $\left|V\right|\le 1$ is satisfied everywhere  away from the RT surface, it will be a valid bit thread configuration, at least in the patch of the manifold that is covered by the coordinates. 

Working in pure AdS we have shown that $\gamma$ depends on $w$ as in eq. \eqref{gammaG}. This will hold for any choice of RT surface. Then, the metric determinant can be written as 
\begin{align}
    \det \gamma_ \lambda &=\det \gamma_ \lambda (0)\, \det \left( \cosh w_ \lambda + \frac{\gamma_ \lambda^{-1}(0)\partial_{w_ \lambda}\gamma_ \lambda(0)}{2} \sinh w_ \lambda \right)^2\,.
\end{align}
As a consequence we find
\begin{equation}
    \left|V\right|= \frac{1}{ \prod_i  \left( \cosh w_ \lambda + \frac{1}{2}D_{ii} \sinh w_ \lambda \right)  }\,,
\end{equation}
where $D$ is the matrix that diagonalizes $\gamma^{-1}_ \lambda\partial_w \gamma_ \lambda$, defined just before equation \eqref{eq:dgamma_gamma_z}. From this equation, we can obtain the constraint
\begin{equation}
\label{eq:bound_dgamma}
    \left|D_{ii}\right|<2 \quad \forall i\in \left\{ 1,\dots, d-1 \right\}\,.
\end{equation}
If this condition is violated there will be some $w$ for which $\det \gamma=0$, making the inverse metric ill-defined. This means the coordinates will cease to be valid at some finite $w$, at which $\left|V\right|\to\infty$, thus, violating the norm bound. This happens when normal geodesics emanating from different points of the RT surface intersect each other. On the other hand, if this condition is satisfied, the coordinates will be well-defined for all $w\in \mathbb{R}$. In other words, all orbits of $\partial_w$ (i.e., all threads) will successfully reach the boundary of the manifold as $w\to\pm\infty$, without intersecting other nearby geodesics. However, this does not guarantee that $\left|V\right|\leq1$ everywhere. To see if this is the case, note that eq. \eqref{eq:condition_dwgamma} must hold true for any RT surface. This implies that at $w_ \lambda=0$ we have
\begin{equation}\label{eq:condw0}
    \Tr \gamma_ \lambda^{-1}\partial_w \gamma_ \lambda = \sum_i D_{ii}=0\,.
\end{equation}
To determine if the norm bound is satisfied or not we can examine the possible extrema of $\frac{1}{\left|V\right|}$. A short calculation shows that
\begin{equation}
    \frac{\dd  }{\dd w_ \lambda}\left( \frac{1}{\left|V\right|} \right)= \frac{1}{\left|V\right|} \sum_ i \frac{\sinh w_ \lambda + \frac{D_{ii}}{2}\cosh w_ \lambda}{\cosh w_ \lambda + \frac{D_{ii}}{2} \sinh w_ \lambda} = \frac{1}{\left|V\right|} \sum_i \tanh(w_ \lambda+ a_i)\,,
\end{equation}
with $a_i\equiv \text{arctanh} \frac{D_{ii}}{2}$. The condition \eqref{eq:condw0} tells us that the sum of hyperbolic tangents vanishes for $w_ \lambda=0$. Since the hyperbolic tangent is monotonic, there are no other values of $w_ \lambda$ that could make the derivative vanish. As a consequence $\left|V\right|$ has only one extremum at $w_ \lambda=0$ and it has to be a global maximum since $\left|V\right|\to 0$ when $w_ \lambda\to \pm \infty$. Thus, $\left|V\right|\le 1$ is satisfied everywhere, provided $\left|D_{ii}\right|<2$ $\,(\forall\,i)$.
Finally, while the coordinates are defined for all $w\in \mathbb{R}$, this does not imply they cover the whole manifold. If that is the case we can define $V=0$ outside the patch covered by the Gaussian normal coordinates and define a piecewise vector field. Such configurations are in fact allowed by the optimization program.

In conclusion, we have found an if and only if condition for the geodesic method to give a valid bit thread configuration in pure AdS, for arbitrary shape deformations. Our proof shows that \eqref{eq:bit_threads_gaussian_normal} is a valid bit thread configuration as long as the eigenvalues of $\gamma^{-1}_ \lambda\partial_w \gamma_ \lambda$ are bounded by $2$ in absolute value or, equivalently, when the normal geodesics do not intersect.\footnote{When the geodesics do intersect, one may be able to untangle them using the procedure outlined in \cite{Headrick:2020gyq}. The construction via integral lines \cite{Agon:2018lwq} would then require a recomputation of the norm. The resulting vector field would be a valid solution, though, one that no longer uses geodesics as integral lines.} It is important to note that this construction only applies when using a single patch to describe the full RT surface. For instance, we cannot use different coordinates for different patches of the same RT surface, nor tackle a more complicated situation with multiple disconnected components of the RT surface, as the ones appearing when studying phase transitions of entanglement entropy or mutual information. Such situations would require different methods to those developed in the present work.

\subsubsection{Geodesic vs. magnetic bit threads}

Armed with the perturbative expansion of the RT surface and the general expression we have found for geodesic bit threads, it is immediate to obtain a perturbative expansion for geodesic bit threads. From (\ref{eq:bit_threads_gaussian_normal}) we find that the result to the second order in $\lambda$ for a general perturbation around a sphere is given by
\begin{align}\label{eq:generalV}
    V\!= &\,\frac{1}{\cosh^{d-1} w} \dd w +  \frac{\lambda}{\cosh^{d-1} w}\left( (d-1) h \tanh w \dd w - \nabla_i h \dd y^{i}  \right) \nonumber\\
    \phantom{=}&\,+ \frac{\lambda^2 }{2 \cosh ^{d-1}w} \bigg\{ (d-1)h^{(1)}\tanh w \dd w -\nabla_i h^{(1)} \dd y^i - \frac{1}{2} \tanh w \nabla_i \left(\nabla^2 h^2 \right) \dd y^i \\
	\phantom{=}&\,+ \bigg[\frac{d}{2}\nabla^2 h^2\tanh^2 w -\frac{1}{2}\nabla^2 h^2+ \nabla_i \nabla_j h \nabla^i \nabla^j h\tanh^2 w  - (d-1)h^2 \tanh^2 w    \bigg] \dd w \bigg\} + \mathcal{O}(\lambda^3).\nonumber
\end{align}
Higher orders are equally straightforward to compute, provided we know $w_ \lambda $ and $\gamma_{ij}^\lambda$ which can be obtained from the perturbations to the RT surface. 
\begin{figure}
    \centering
    \includegraphics[width=15.8cm]{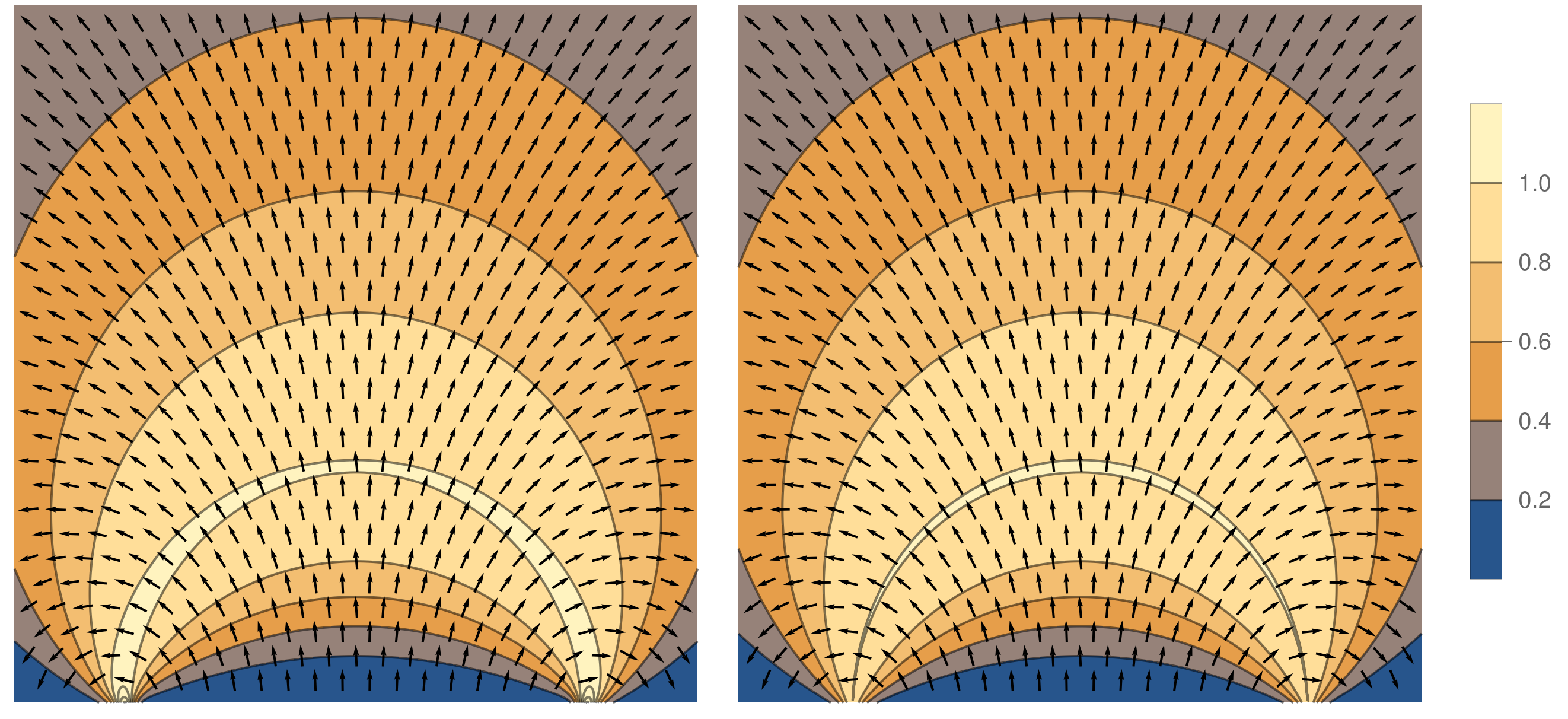} 
    \caption{Contour plots for the modulus of the bit threads of an ellipse with $\lambda=\varepsilon^2=0.1$ up to the first order in $\lambda$. A this order, the solution for $V$ based on the Gaussian method (\ref{eq:generalV}) is in agreement with the Biot-Savart law. The left figure shows the $y=0$ plane, a cross-section of the ellipse containing the major axis. The right figure shows the $x=0$ plane, which contains the minor axis. We take the semi-minor axis to be the radius of the original disk. A slight violation of the norm bound is induced by the truncation. It can be shown that such violation is of order $\mathcal{O}(\lambda^2)$, and is cured by including higher orders in the expansion.
    \label{fig:bit_thread_perturbations_contour}}
\end{figure}
As a concrete example, we plot the result for the case of the ellipse in Figure \ref{fig:bit_thread_perturbations_contour}. It is clear from this plot that truncating the perturbative series leads to a slight violation of the norm bound. This is not a problem and, in fact, one can check that including higher-order terms in the expansion reduce the apparent violations. This is equivalent to the behavior observed when studying small perturbations of the CFT state \cite{Agon:2020mvu}. Ultimately, the norm bound will be satisfied as long as we include the whole perturbative series, as we are supposed to do. 

It is interesting to compare the perturbations obtained here with those coming from the magnetic formula. Remarkably, we find that at least up to the first order in $\lambda$ they match perfectly. At the second order, they still match, but only on the RT surface. This indicates that magnetic bit threads reproduce the right entanglement entropy up to the second order but they will not follow geodesics. As for higher-order perturbations, we know for sure that the magnetic formula cannot reproduce the right result beyond the second order, since at that point the functions $F_n$ start playing a role and the nonlocality of entanglement entropy manifests itself. However, it seems that the magnetic formula completely accounts for the local contributions to the bit threads. Then, as long as we are only interested in small perturbations, we can still use the much simpler magnetic prescription, bearing in mind that the approximation will eventually break down.

\section{Conclusions and outlook}
\label{sec:conclusion}

The conclusion of the previous section is also the main conclusion of our paper: bit threads, hence holographic entanglement, can be described as magnetic field lines in the bulk which are sourced by a hypothetical electric current that flows along the boundary of the entangling surface in the dual field theory; see Fig. \ref{Vhalfplane} for an illustration. Even though this description is strictly valid when the entangling surface --- hence the corresponding RT surface --- is maximally symmetric, we have seen that ``magnetic'' bit threads continue to match the ``geodesic'' ones at the leading order in deformation of the entangling surface. 

For larger deformations nonlocal corrections eventually become inevitable and description in terms of local magnetic field lines should be replaced with some sort of nonlocal magnetism, in line with the nonlocality of the modular hamiltonian for generic entangling surfaces. Instead of trying to figure out this nonlocal theory, in section \ref{sec:deform} we developed a perturbative calculation scheme using Gaussian normal coordinates to compute corrections to the RT surface and corresponding geodesic bit threads. Quite interestingly, we showed that such an expansion admits a natural interpretation in terms of Witten diagrams. This scheme allows for computing corrections to the entanglement entropy works to arbitrary order in the deformation and is more straightforward to implement than the previous methods in the literature. We also concluded that the description of bit threads in terms of geodesics in the bulk is quite generic and continues to be a valid description including all nonlocal corrections, as long as they do not intersect. 

We have not tried to extend our methods to explore situations where geodesics intersect. This will happen generically for a class of large deformations (e.g., for concave entangling regions) and for cases with more complicated topologies such as multi-component surfaces. As discussed in subsection \ref{sec:defgeo}, magnetic bit threads do not follow geodesics beyond the leading order in the deformation. It should be interesting to directly consider non-local modifications to the Biot-Savart law based on our diagramatic expansion and apply such formalism to the cases where the geodesic construction breaks down. This may shed light on questions regarding phase transitions of entanglement entropy and mutual information. 

Our work can be expanded in several other interesting directions. We confined our analysis in section \ref{sec:magnetic} to entanglement in $(2+1)$-dimensional conformal field theories. Generalization to higher dimensions should be straightforward, upon upgrading the electric currents to surface current densities. Moreover, our perturbative scheme in section \ref{sec:deform} to deal with shape deformations is already valid for arbitrary $d$; it should be worth exploring some concrete examples in higher dimensions and compare with previous numerical results. Generalization to non-CFTs seems less straightforward --- but equally interesting --- because we have used the maximal symmetry of the bulk AdS numerous times both in sections \ref{sec:magnetic} and \ref{sec:deform}. Another interesting question concerns the generalization of our setup beyond the vacuum state. Of particular interest would be to consider thermal states, which are dual to black hole or black brane geometries. One should also explore the possibility of extending our classical bit threads to quantum ones. As argued in \cite{Agon:2021tia,Rolph:2021hgz}, quantum bit threads require a non-vanishing divergence of the corresponding vector field. In the picture presented in this paper, this quite generically seems to require the presence of magnetic monopoles in the bulk.  A final consideration would be to develop a similar perturbative scheme and a magnetic-like interpretation for Lorentzian threads, extending \cite{Pedraza:2021mkh,Pedraza:2021fgp,Pedraza:2022dqi}. This would enable us to better understand the nature of holographic complexity and its response under shape deformations of the boundary Cauchy slice. We leave these interesting questions for future investigation. 

\section*{Acknowledgments}

It is a pleasure to thank Aylon Pinto for his collaboration in the initial stages of this project. We also thank C\'esar Ag\'on, Rafael Carrasco, Casey Cartwright, Veronika Hubeny, Andrew Svesko and Erik Tonni for useful discussions and comments on the manuscript. UG is supported by the Netherlands Organisation for Scientific Research (NWO) under the VICI grant VI.C.202.104. JFP is supported by the `Atracci\'on de Talento' program (2020-T1/TIC-20495, Comunidad de Madrid) and by the Spanish Research Agency (Agencia Estatal de Investigaci\'on) through the grants CEX2020-001007-S and PID2021-123017NB-I00, funded by MCIN/AEI/10.13039/501100011033 and by ERDF A way of making Europe.

\appendix

%%%%%%%%%%%%%%%%%%%%%%%%%%%%%%%%%%%%%%%%%%%%%
%%%%%%%%%%%%%%%%%%%%%%%%%%%%%%%%%%%%%%%%%%%%%
\section{Flux maximization of the modified Biot-Savart law}
\label{Fluxmaximization}
%%%%%%%%%%%%%%%%%%%%%%%%%%%%%%%%%%%%%%%%%%%%%
%%%%%%%%%%%%%%%%%%%%%%%%%%%%%%%%%%%%%%%%%%%%%

To show that expression \eqref{eq:magneticfield} is actually a local maximum of the flux functional we have to show that it fulfills equation \eqref{eq:modifiedMaxwellRT}. We can explicitly compute
\begin{align}
f\nabla \times \boldsymbol{B}&=- \frac{2f}{\pi} \nabla \times \int \dd s \;\frac{(\boldsymbol{r}-\boldsymbol{r}_0(s))}{\left|\boldsymbol{r}- \boldsymbol{r}_0(s)\right|^4}\times \frac{\dd \boldsymbol{r}_0}{\dd s} \nonumber\\
&= \frac{2}{\pi} \int \dd s\; \frac{\dd \boldsymbol{r}_0}{\dd s} \frac{-2}{\left|\boldsymbol{r}- \boldsymbol{r}_0(s)\right|^4}  + (\boldsymbol{r}- \boldsymbol{r}_0(s))\frac{\dd }{\dd s}  \frac{1}{\left|\boldsymbol{r}- \boldsymbol{r}_0(s)\right|^4}\,,
\end{align}
and for the other term in equation \eqref{eq:modifiedMaxwellRT}
\begin{align}
\nabla  f \times \boldsymbol{B}&=- \frac{2f}{\pi}\int \dd s \; \frac{\nabla \ln f\times (\boldsymbol{r}- \boldsymbol{r}_0(s))\times \frac{\dd \boldsymbol{r}_0}{\dd s} }{\left|\boldsymbol{r}- \boldsymbol{r}_0(s)\right|^4} \nonumber\\
&=- \frac{2f}{\pi} \int \dd s\; \frac{\boldsymbol{r}- \boldsymbol{r}_0(s)}{\left|\boldsymbol{r}- \boldsymbol{r}_0(s)\right|^4} \left( \nabla \ln f \cdot \frac{\dd \boldsymbol{r}_0}{\dd s}  \right) - \frac{\dd \boldsymbol{r}_0}{\dd s} \left( \nabla \ln f \cdot  \frac{\boldsymbol{r}- \boldsymbol{r}_0(s)}{\left|\boldsymbol{r}- \boldsymbol{r}_0(s)\right|^4} \right)\,.
\end{align}
Putting both of them together we obtain
\begin{align}
\nabla \times(f \boldsymbol{B})& = \frac{2f}{\pi} \int \dd s\; \frac{\dd \boldsymbol{r}_0}{\dd s} \frac{\left( \nabla \ln f \cdot (\boldsymbol{r}- \boldsymbol{r}_0(s))-1 \right) }{\left|\boldsymbol{r}- \boldsymbol{r}_0(s)\right|^4} - \frac{\boldsymbol{r}- \boldsymbol{r}_0(s)}{\left|\boldsymbol{r}- \boldsymbol{r}_0(s)\right|^4} \frac{\dd \boldsymbol{r}_0}{\dd s}\cdot  \nabla \ln f\,,
 \end{align}
where we have dropped a total derivative which vanishes due to the fact that $\boldsymbol{r}(s)$ is a closed curve. By using $f(x,y,z)=Kz$ for some constant $K$ we immediately find 
 \begin{equation}
 \label{eq:modifiedmaxwelsolution}
     \nabla \times \left( K z \boldsymbol{B} \right)=0\,,
 \end{equation}
given that we only use currents localized on the boundary so that $\boldsymbol{r}_0(s)$ has no $z$ component. Note that this only applies when $\boldsymbol{r}\neq \boldsymbol{r}_0(s)$ for any $s$ so that the integral is convergent. To see what happens when we get close to the sources let us pick an arbitrary point on the entangling surface. Then, consider the surface integral of the left-hand side of this equation on a small semi-circle of radius $\delta$ normal to the local direction of the current. Without loss of generality, we can assume that $\boldsymbol{r}_0(s)\approx  \epsilon \hat{\boldsymbol{z}}+ J \hat{\boldsymbol{y}} s$ on the vicinity of this point. Note that we are regulating the entangling surface by putting it on the $z=\epsilon$ plane instead of the actual conformal boundary. This is necessary to compute the integral and only leads to order $\epsilon$ violations to equation \eqref{eq:modifiedmaxwelsolution} which disappear when the regulator is removed. Then we find
\begin{equation}
    \int \nabla \times (K z \boldsymbol{B})\cdot \dd S=K\int z \boldsymbol{B}\cdot \dd \boldsymbol{\ell}=-  \frac{2 K}{\pi} \int \dd s \dd \theta \frac{J\delta^3\sin \theta}{(\delta^2 + J^2 s^2)}= - 2K \text{sign}(J)\,,
\end{equation}
keeping only the terms up to $\mathcal{O}(\epsilon^0)$. Since this integral does not vanish the original curl must be proportional to a Dirac delta. The only possibility which will reproduce this result close to all points in the entangling surface is
\begin{equation}
    \nabla \times(f B)=- 2 K \int \dd s \delta^{(3)}(\boldsymbol{r}- \boldsymbol{r}_0(s)) \frac{\dd \boldsymbol{r}_0}{\dd s}= 2 K \boldsymbol{J}\,,
\end{equation}
which tells us that $K=\frac{1}{2}$. Then, the modified magnetic field is a local maximum of the flux and fulfills the equation not only on the RT surface but everywhere in the bulk.

%%%%%%%%%%%%%%%%%%%%%%%%%%%%%%%%%%%%%%%%%%%%%
%%%%%%%%%%%%%%%%%%%%%%%%%%%%%%%%%%%%%%%%%%%%%
\section{RT surface perturbation for the ellipse}
\label{ellipse_computation}
%%%%%%%%%%%%%%%%%%%%%%%%%%%%%%%%%%%%%%%%%%%%%
%%%%%%%%%%%%%%%%%%%%%%%%%%%%%%%%%%%%%%%%%%%%%

Given the equation for $h$ \eqref{eq:h_ellipse} we can directly compute $F_2(r_s,\theta)$. The next step is to expand it in Fourier modes and solve equation \eqref{eq:non-homogeneous_part} for each of the Fourier modes. Overall one finds
\begin{align}
    g&=\frac{-b r_s^2 (17 \sqrt{b^2-r_s^2}+15 b)+4 b^3 (5
   \sqrt{b^2-r_s^2}+3 b)+3 r_s^4}{32 \sqrt{b^2-r_s^2}
  (\sqrt{b^2-r_s^2}+b)^3} \nonumber\\
   & \phantom{=}+\left[\frac{28 b^7 (\sqrt{b^2-r_s^2}-b)+b^5 r_s^2 (103 b-89
   \sqrt{b^2-r_s^2})}{10 r_s^8} \right.\nonumber\\
   &\phantom{=}+ \left.\frac{2 b r_s^6 (1975 b-891 \sqrt{b^2-r_s^2})+32 b^3 r_s^4
   (173 \sqrt{b^2-r_s^2}-255 b)-485 r_s^8}{640 r_s^8}\right]\cos 2 \theta \\
   &\phantom{=}+\frac{3 r_s^2 (b (\sqrt{b^2-r_s^2}-b)+r_s^2)}{32
   \sqrt{b^2-r_s^2} (\sqrt{b^2-r_s^2}+b)^3}\cos 4 \theta \nonumber\\
   &\phantom{=}-\frac{3 [6 b r_s^4 (\sqrt{b^2-r_s^2}-3 b)+32 b^5
   (\sqrt{b^2-r_s^2}-b)+16 b^3 r_s^2 (3 b-2
   \sqrt{b^2-r_s^2})+r_s^6]}{128 r_s^6} \cos 6 \theta\,,\nonumber
\end{align}
We have chosen the coefficients such that we keep all the divergences in the homogeneous part of the solution, thus this function is finite both when $r_s\to 0$ and when $r_s\to b$. Note, however, that its derivatives do have divergences.
With this solution in mind one can solve equation \eqref{eq:z=epsilon} up to third order in $\lambda$ keeping the coefficients in $h^{(1)}$ and $h^{(2)}$ generic. This gives us an expression for the value of $r_s=r_s|_{z=\epsilon}$ where the RT surface intersects the surface $z=\epsilon$ as a function of $b$, $\epsilon$, $\theta$ and the coefficients yet to be determined. One must then use this solution to expand $r(r_s|_{z=\epsilon}
,w|_{RT}(r_s|_{z=\epsilon}),\theta)$ up to third order. Imposing that the result equal the boundary conditions of equation \eqref{eq:boundary_conditions_ellipse} one finds for the coefficients
\begin{align}
     h^{(1)}&= \frac{3 b}{16} f_0(r_s) + \frac{b}{4} f_2(r_s) \cos 2 \theta + \frac{b}{16}  f_4(r_s) \cos 6 \theta\,, \\
     h^{(2)}&=\left(\frac{99b}{256}+\frac{15 b ^3}{64 \epsilon^2}\right)f_0(r_s) + \left(\frac{609 b}{512}+ \frac{45 b^3}{128 \epsilon^2}\right) f_2(r_s) \cos 2 \theta \nonumber\\
     &\phantom{=}+\left(\frac{309b}{256}+ \frac{9b^3}{64 \epsilon^2}\right)f_4(r_s) \cos 4 \theta +\left(\frac{207b}{512}+ \frac{3b^3}{128 \epsilon^2}\right)f_6(r_s) \cos 6 \theta+g\,.
 \end{align}
It is important that some of the $h^{(2)}$ coefficients are divergent in $\epsilon$, this fact has to be taken into account when solving \eqref{eq:z=epsilon} to obtain a consistent result. All that is left is to integrate the area element from $r_s=0$ up to $r_s|_{z=\epsilon}$. A final $\theta$ integration must also be performed. The result is as quoted in the main text, in equation \eqref{eq:ent_entr_ellipse}.

\bibliographystyle{JHEP}
\bibliography{refs-SBT}

\end{document}